\PassOptionsToPackage{sort&compress}{natbib}
\documentclass[preprint,12pt]{elsarticle}

\usepackage{amssymb}
\usepackage{amsmath}
\usepackage{chemformula} 
\usepackage[T1]{fontenc} 
\usepackage[version=4]{mhchem}
\usepackage{enumerate}
\usepackage{pdflscape}
\usepackage{multicol}
\usepackage{array}
\usepackage{dcolumn}
\usepackage{multirow}
\usepackage{url}
\usepackage[symbol]{footmisc}
\usepackage{threeparttable}
\usepackage{booktabs}
\usepackage{afterpage}
\usepackage{xcolor}
\usepackage{soul}
\usepackage{blindtext}
\usepackage{longtable}
\usepackage{hhline}
\usepackage{caption}
\usepackage{subcaption}
\usepackage{graphicx}
\usepackage{amsmath}
\usepackage{footnote}

\journal{JQSRT}

\begin{document}

\begin{frontmatter}


\title{Experimental and theoretical investigation on \ce{N2} pressure-induced coefficients of the lowest rotational transitions of \ce{HCN}} 

\author{Francesca Tonolo\corref{cor1}\fnref{rennes}}
\ead{francesca.tonolo.1@univ-rennes.fr}
\author{Hubert Jóźwiak\fnref{torun}}
\author{Luca Bizzocchi\corref{cor1}\fnref{ciamician}}
\ead{luca.bizzocchi@unibo.it}
\author{Mattia Melosso \fnref{ciamician}}
\author{Piotr Wcisło\fnref{torun}}
\author{François Lique\fnref{rennes}}
\author{Cristina Puzzarini\fnref{ciamician}}

\cortext[cor1]{Corresponding authors}

\affiliation[rennes]{organization={Univ. Rennes, CNRS, IPR (Institut de Physique de Rennes)},
             addressline={UMR 6251},
             city={Rennes},
             postcode={F-35000},
             country={France}} 
\affiliation[torun]{organization={Institute of Physics, Faculty of Physics, Astronomy and Informatics, Nicolaus Copernicus University in Toruń},
             addressline={Grudziadzka 5, 87-100},
             city={Toruń},
             country={Poland}}              
\affiliation[ciamician]{organization={Dipartimento di Chimica ``Giacomo Ciamician'', Università di Bologna},
             addressline={Via F.~Selmi 2},
             city={Bologna},
             postcode={40126},
             country={Italy}} 

\begin{abstract}
We present the first experimental determination of room-temperature \ce{N2} pressure broadening, speed dependent broadening, and pressure shift coefficients of the three lowest rotational lines of \ce{HCN}. The experimental results served to assess the accuracy of a low-cost yet accurate computational strategy, which relies on a simplified characterization of the \ce{HCN-N2} interaction potential, and employs a novel approximate method of solving the quantum scattering problem. 
Building on the validation of this computational approach, the dataset was extended to higher rotational transitions, up to $J(\ce{HCN})=5\leftarrow4$. 
For these transitions, 
we provide the temperature dependence of the pressure broadening coefficient, its speed dependence parameter, and the Dicke narrowing parameter. This new dataset can support and refine the modeling of \ce{HCN} in both the terrestrial and Titan's atmospheres. This work constitutes an important step towards populating spectroscopic databases with accurate \ce{HCN} line-shape parameters.

\end{abstract}





\end{frontmatter}

\section{Introduction}
\label{sec1}

Hydrogen cyanide (\ce{HCN}) is a pivotal tracer of chemical networks in terrestrial and extraterrestrial environments, where it is found in significant abundances.
In the Earth's atmosphere, \ce{HCN} serves as a reliable marker of biomass burning emissions~\citep{lobert1990importance,holzinger1999biomass} and plays a crucial role in the nitrogen cycle and reactive nitrogen chemistry~\citep{li2000atmospheric}. Beyond Earth, \ce{HCN} is one of the most important proxy for chemical evolution, being detected in high abundances across diverse interstellar environments, such as molecular clouds~\citep{snyder1971observations,hirota1998abundances,liszt2001comparative}, star-forming regions, and circumstellar envelopes~\citep{rice2018exploring,thi2004organic,oberg2021molecules}. In addition, it has been identified in comets~\citep{rodgers1998hnc,huebner1974hcn,wirstrom2016hcn,cordiner2019alma,cordiner2023gas}, in many planetary atmospheres, e.g., the ones of Jupiter, Neptune, Uranus, Pluto, the super-Earth 55\,Cancri, and in Titan, the largest moon of Saturn~\citep{tokunaga1981detection,marten1993first,molter2016alma,tsiaras2016detection,lellouch2017detection,rengel2022ground}.


Particular attention should be given to the role of hydrogen cyanide in Titan atmosphere, where it exhibits vertical gradients from its primary formation site in the upper atmosphere to condensation below 80\,km of altitude~\citep{yelle1991non,thelen2018spatial}. Notably, emissions from \ce{HCN} rotational states are among the dominant thermospheric cooling processes on Titan, helping to balance heating of the ultraviolet haze and thermal conduction~\citep{rezac2013rotational,thelen2022variability}. However, the local thermodynamic equilibrium (LTE) approximation, commonly used to infer the population distribution of rotational states, breaks down above approximately 1100\,km, leading to inaccuracies in cooling rate predictions~\citep{rezac2013rotational}.
These discrepancies are primarily due to the collisional dynamics with \ce{N2}, the most abundant constituent of Titan atmosphere. In fact, state-to-state population transfer and pressure broadening effects vary significantly with altitude, thus producing a continuum of emission intensities and line shapes~\citep{thelen2022variability}. 
Moreover, understanding pressure-induced effects on spectral line shapes is also essential for accurately modeling the Earth's atmospheric composition, where nitrogen (\ce{N2}) is the primary perturber. These effects significantly influence radiative transfer calculations, also affecting the precision of climate and weather forecasts~\cite{pine1993self,gordon2022hitran2020}.

Generally speaking, the accurate knowledge of pressure-induced effects on \ce{HCN} spectroscopic lines is crucial for atmospheric monitoring, climate evolution simulations, and radiative transfer modeling in both terrestrial and astrophysical environments. 
Room temperature collisional broadening of \ce{HCN} lines due to \ce{O2}, \ce{N2}, and air
has been extensively studied in the last few decades~\citep{Varghese_1984,pine1993self,D_Eu_2002, rinsland2003multispectrum,devi2004multispectrum,devi2005multispectrum,smith2008low,yang2008oxygen,Rohart_2021}, and a substantial amount of data has already been incorporated into the HITRAN database~\citep{gordon2022hitran2020}. 
Currently, such compilation includes \ce{N2}- and air-pressure broadening coefficients for many \ce{HCN} transitions; they were computed using a polynomial function whose coefficients were derived from a fit to all available experimental data, assuming no dependence on vibrational state or isotopic species.
However, due to the lack of measured \ce{N2}-broadening coefficients for rotational transitions with $J<5$, the lower-energy end of this extrapolation remains unverified.

It has to be noted that the lowest rotational transitions of \ce{HCN} exhibit the hyperfine splitting due to the nuclear qua\-dru\-pole coupling of the \ce{^{14}N} nucleus ($I = 1$), at least, partially resolved, and their analysis thus requires particular attention.
To circumvent the complications involved in the analysis of the hyperfine structure, \citet{colmont1985collisional,rohart2007lineshapes} and~\citet{Rohart_2021} investigated the \ce{N2}-broadening of the R(0), R(1) and R(2) transitions\footnote{Throughout this work, we adopt the R($J$) notation, where R($J$) refers to the $J+1 \leftarrow J$ rotational transition.} of \ce{HC^{15}N} isotopologue, that shows no quadrupole coupling
due to the $I = 1/2$ spin of $^{15}$N.
An estimate of the \ce{N2} broadening coefficient of the R(3) transition of \ce{HC^{14}N} (recorded at 304\,K) is reported in the Appendix of Ref.~\citep{rohart2007lineshapes}, where the line profile fit was performed assuming the same collisional width for all the hyperfine components. 



\ce{HCN} line shapes exhibit pressure-induced effects that extend beyond the standard Voigt profile, including the speed-dependence of collisional broadening~\citep{Berman_1972}. This effect has been studied in selected ro-vibrational transitions of \ce{HCN}~\citep{D_Eu_2002, smith2008low, rohart2007lineshapes, Rohart_2021}, assuming a quadratic dependence of the collisional width on the speed of the absorbing molecule (see the next section for details)~\citep{Rohart1994}. However, a comprehensive study of this effect in \ce{HCN} is still lacking, especially for pure rotational transitions.
Additionally, while the HITRAN database has gradually incorporated quadratic speed-dependence coefficients for other molecules~\cite{Wcislo_2016, Stolarczyk_2020}, it still does not provide these parameters for \ce{HCN} lines. 
In addition to speed-dependent collisional broadening, several experimental studies of \ce{HCN} lines~\citep{Varghese_1984, pine1993self, D_Eu_2002, rohart2007lineshapes, Rohart_2021} have attempted to quantify line narrowing caused by velocity-changing collisions~\citep{dicke1953}, but the retrieved narrowing parameters often led to ambiguous physical interpretations (see Ref.~\citep{Rohart_2021}). 

Another limitation of the HITRAN database on \ce{HCN} is the lack of information about the temperature dependence of the pressure broadening coefficients. This is particularly significant for observations of the Titan’s atmosphere, where temperatures remain below $\sim180$\,K and low-$J$ rotational transitions are extensively used to probe the chemical and dynamical processes (see, e.g.,~Refs.~\citep{hidayat1997millimeter,marten2002new,lellouch2019intense,thelen2022variability}).  Conversely, extending broadening data to higher temperatures (above 300\,K) is equally crucial for terrestrial atmospheric studies, where accurate modeling is needed for climate and remote sensing applications.

In this work, we address all these lacks through a combined experimental and theoretical study of the first three rotational transitions [R(0), R(1) and R(2)] of \ce{HCN} perturbed by \ce{N2}. We performed high-precision measurements of these lines at 296~K, determining both the pressure broadening coefficients and their speed-dependence parameters. The experimental results were employed to validate \textit{ab initio} calculations of these line shape parameters obtained by solving the quantum scattering problem using a novel, computationally feasible approach and an improved \ce{HCN}-\ce{N2} potential energy surface (PES). 
The calculations employed spin-free scattering matrices, neglecting hyperfine coupling effects in the retrieval of line shape parameters, as recoupling calculations fall beyond the primary scope of the work.
This validation permitted to extend the calculations across a broad temperature range ($100-800$~K), thus providing the temperature dependence of the pressure broadening, its speed-dependence, and line narrowing coefficients for the R(0)--R(4) transitions. 

This work is organized as follows. 
In \S~\ref{sec2}, the experimental setup and the methodology employed to record and fit the line profiles, within the determination of the corresponding pressure-induced coefficients, are outlined. 
\S~\ref{sec3} discusses the computational strategy employed for the characterization of the collisional PES as well as the open channels approach for scattering calculations.
Subsequently, the computational-experimental comparison of the results is presented, together with the corresponding extension of the computed values to higher $J$ rotational lines, and their temperature-dependence coefficients (\S~\ref{sec4}). 
In the last section (\S~\ref{sec5}), conclusions are drawn.

\section{Experimental details}
\label{sec2}

\subsection{Experimental setup}
\label{sub2.1}
Pressure broadening experiments were conducted using the frequency-modulated millimeter-/submillimeter-wave (FM-mmW) spectrometer of the Rotational and Computational Spectroscopy Lab at the University of Bologna. 
A detailed description of the experimental setup is provided elsewhere~\citep{melosso2019pure,melosso2019rotational}. Briefly, this instrument employs several Gunn diodes as microwave radiation sources which cover the 75--134\,GHz frequency range; higher frequencies are obtained by passive multipliers. 
The absorption cell consists of a 3\,m long Pyrex tube with a 5\,cm diameter, sealed at either ends with polyethylene windows. 
To improve the signal/noise ratio, sine-wave frequency modulation is applied. The output signal, detected by Schottky barrier diodes, is subsequently demodulated by a Lock-in amplifier tuned at twice the modulation frequency ($2f$). This yields the second derivative of the actual absorption profile. 

The \ce{HCN} sample was prepared from mandelonitrile (\ce{C6H5-CHOH-CN}) vapours. 
In the gas phase, this molecule is fully dissociated into benzaldehyde and \ce{HCN}. To separate these components, benzaldehyde was condensed in a trap kept in a ice slush bath. The purified \ce{HCN} gas was then collected in a second trap held at liquid nitrogen temperature. 
The temperature of this reservoir was then raised to approximately 253\,K, at which \ce{HCN} has a vapor pressure suitable for the measurements.
For each transition, 1.5\,mTorr of \ce{HCN} were first introduced into the cell; then, six series of measurements were carried out at increasing values of the pressure of the \ce{N2} buffer gas (99.99\% pure), ranging from $\sim 4$ to $\sim 220$\,mTorr, with steps of approximately 10\,mTorr. 
Given the sample preparation conditions, no spectroscopic features of benzaldehyde were observed in the cell.
In total, 186, 103 and 110 spectral recordings at room temperature ($\sim 296$\,K) and different pressures of \ce{N2} were carried out for the R(0), R(1) and R(2) transitions, respectively.
The gas pressure was found to be stable well within 0.2\,mTorr. 
The measurements were performed under static conditions.

\begin{figure*}  
\centering
 \includegraphics[scale=0.33]{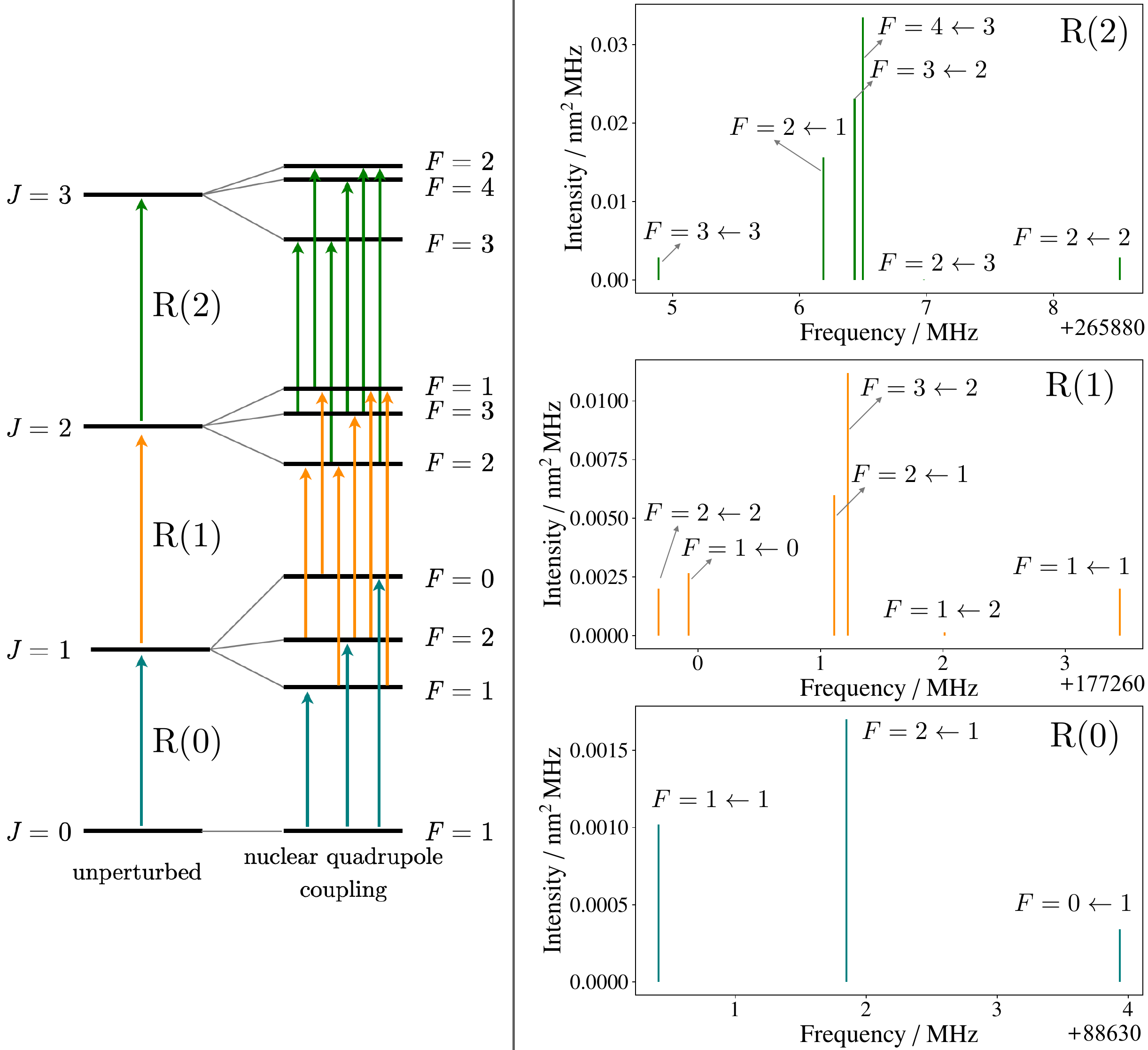}
 \caption{Illustration of the splitting of rotational levels of \ce{HCN} due to nuclear quadrupole coupling. The right panels show the resulting hyperfine structure together with the integrated intensities of each hyperfine transition at 300\,K.}
 \label{fig_intro} 
\end{figure*}

\subsection{Line profile analysis}
\label{sub2.2}
A frequency-modulated, quadratic speed-dependent Voigt profile (\mbox{qSDVP}; \citep{berman1972speed,dore2003using}) was employed to model the line profile of each spectral recording. 
This approach implements the full complex representation of the Fourier-transformed correlation function and uses as fitting parameters: the Lorentzian line half-width ($\Gamma_0$), the quadratic speed-dependent relaxation coefficient ($\Gamma_2$), the shift from the central frequency ($\Delta_0$), and the dispersion factor ($\phi$). For further details, the reader is referred to~Ref.~\cite{dore2003using}.

\begin{figure*}  
\centering
 \includegraphics[scale=0.54]{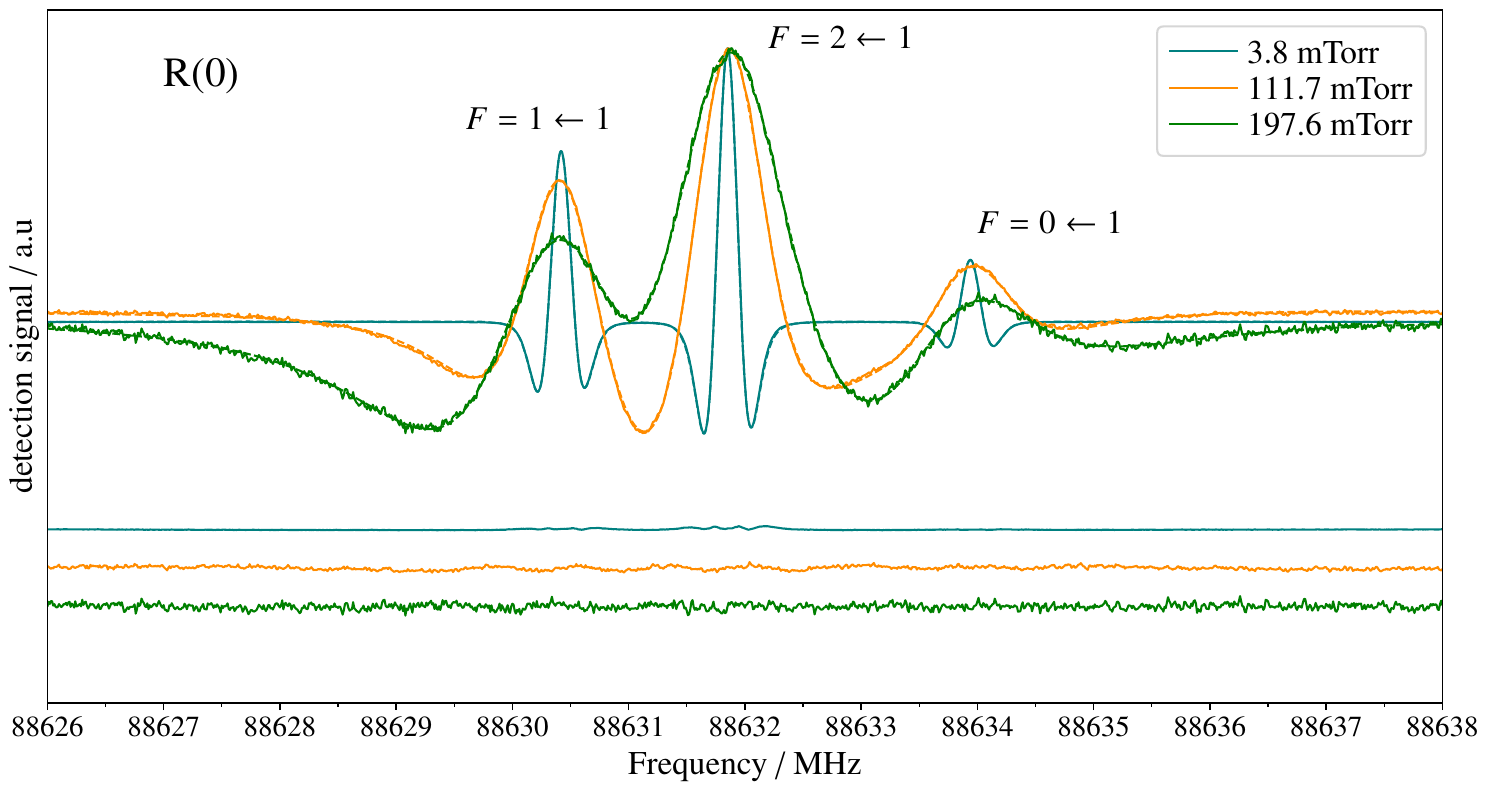}
 \caption{Line profile analysis of three measurements for the R(0) transition of \ce{HCN} at different pressures of the buffer gas of \ce{N2} (the minimum and maximum values recorded, and an  intermediate one). The qSDVP model is employed. At the bottom of the figure, residuals (i.e., the observed-calculated differences) are also reported. $T=296$\,K.}
 \label{fig6} 
\end{figure*}

\begin{figure*}  
\centering
 \includegraphics[scale=0.21]{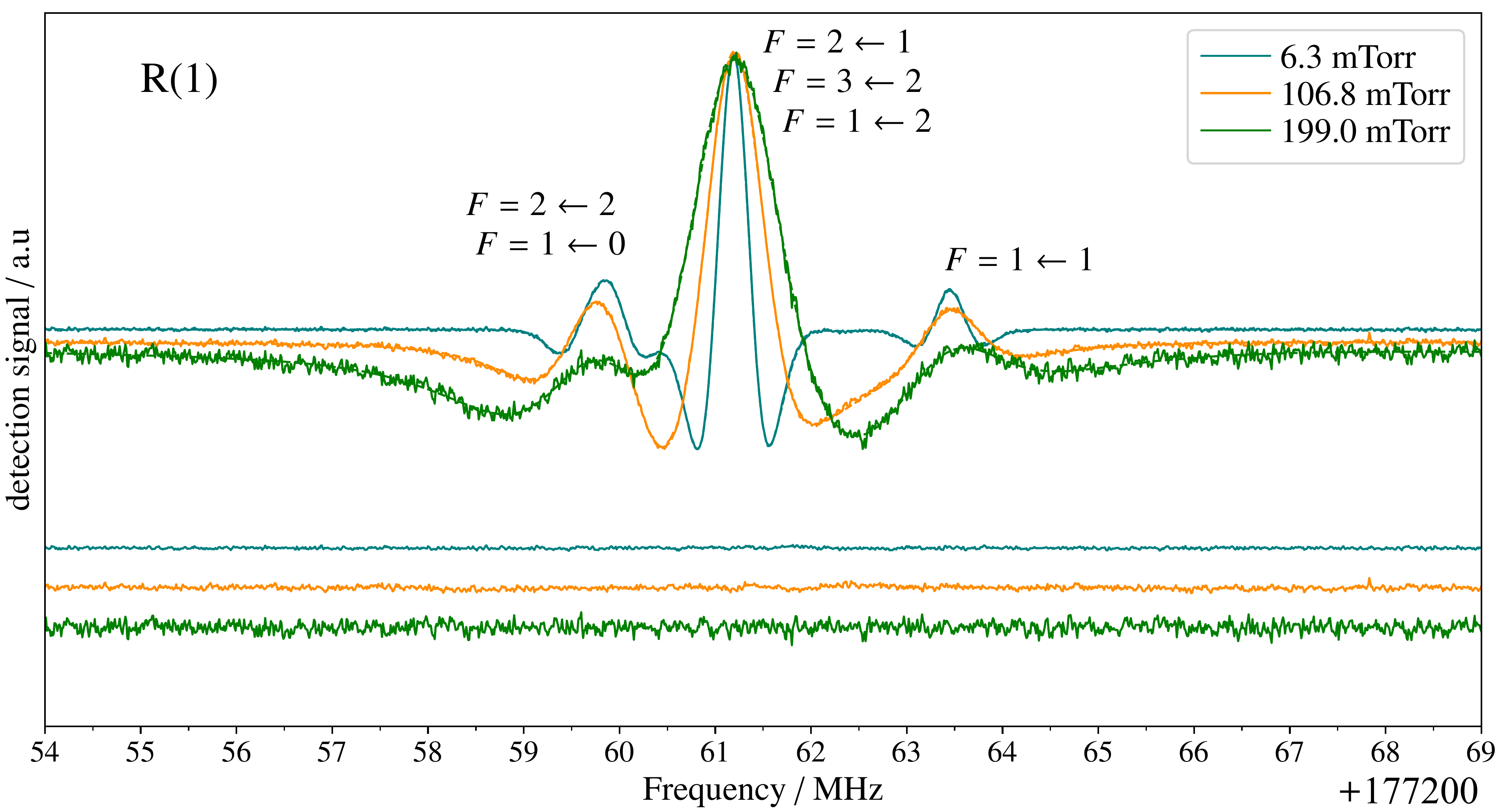}
 \caption{Line profile analysis of three measurements for the R(1) transition of \ce{HCN} at different pressures of the buffer gas of \ce{N2} (the minimum and maximum values recorded, and an  intermediate one). The qSDVP model is employed. At the bottom of the figure, residuals (i.e., the observed-calculated differences) are also reported. $T=296$\,K.}
 \label{fig6a} 
\end{figure*}

\begin{figure*}  
\centering
 \includegraphics[scale=0.21]{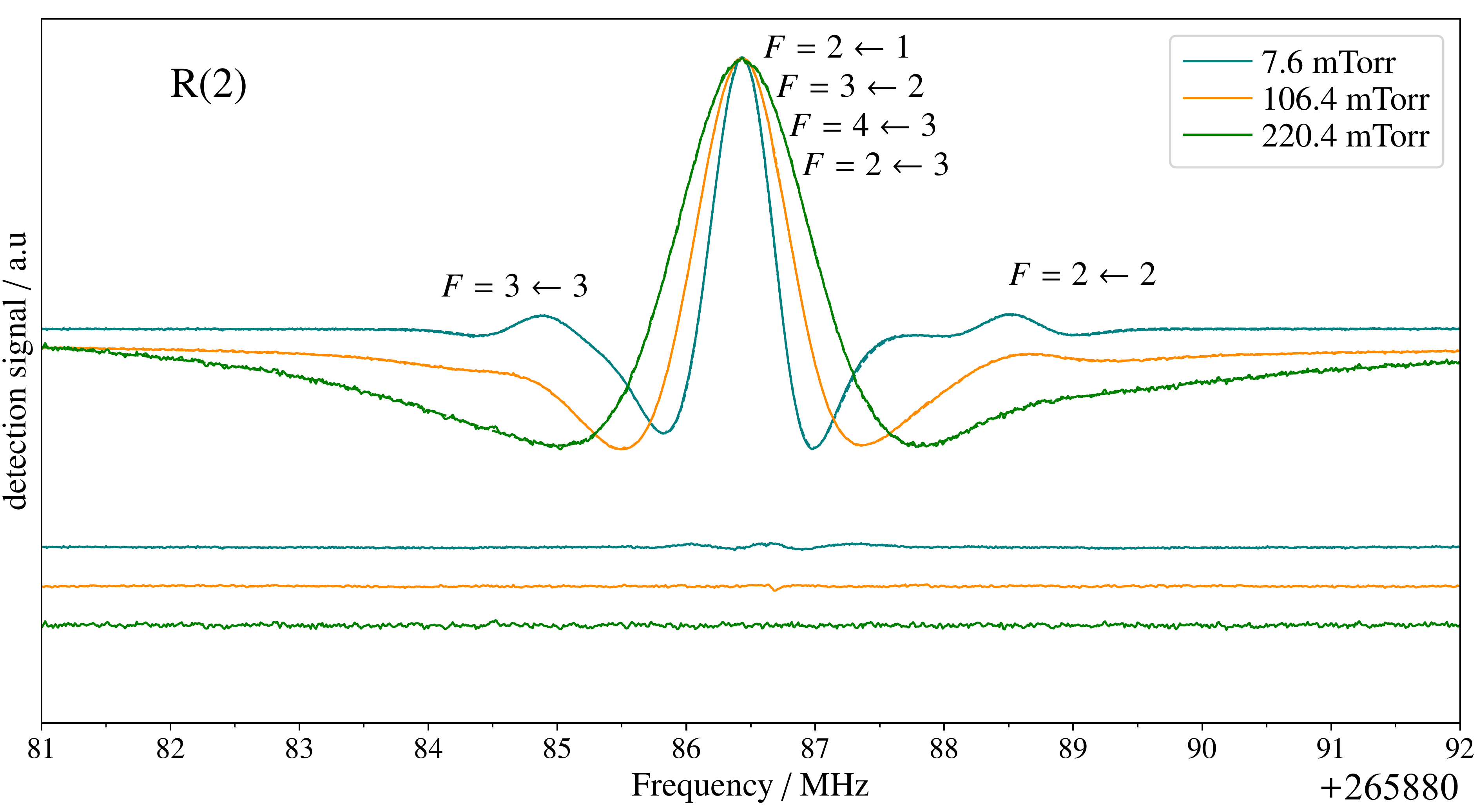}
 \caption{Line profile analysis of three measurements for the R(2) transition of \ce{HCN} at different pressures of the buffer gas of \ce{N2} (the minimum and maximum values recorded, and an  intermediate one). The qSDVP model is employed. At the bottom of the figure, residuals (i.e., the observed-calculated differences) are also reported. $T=296$\,K.}
 \label{fig6b} 
\end{figure*}

As mentioned before, the observed transitions are split due to nuclear quadrupole coupling between the nuclear spin of \ce{^{14}N} ($I=1$) and the molecular end-over-end rotation. This leads to the splitting of each rotational level into three $F$ sublevels, with $F$ varying between $|I-J|$ and $I+J$. 
For illustrative purposes, Fig.~\ref{fig_intro} depicts the splitting scheme of the lowest rotational levels of \ce{HCN} and the resulting hyperfine structure of the relevant rotational transitions.
Few examples of the experimental and modeled traces for R(0), R(1), and R(2) transitions of \ce{HCN} are presented in Figs.~\ref{fig6},~\ref{fig6a}, and~\ref{fig6b}, respectively, for three measurements at different pressures.
Overall, the fits show very small, and almost featureless, residuals across all pressures, thereby supporting the choice of the qSDVP model for the line profile.
The hyperfine structure of the R(0) transition consists of three clearly separated components ($F=1\leftarrow1$, $F=2\leftarrow1$, $F=0\leftarrow1$) as shown in Fig.~\ref{fig6}. 
Conversely, for the R(1) and R(2) transitions, at the instrumental resolution, only three features are discernible. 
For the R(1), the lowest-frequency component corresponds to the overlap of the $F=2\leftarrow2$ and $F=1\leftarrow0$ transitions, the central line results from the overlap of the $F=2\leftarrow1$, $F=3\leftarrow2$ and $F=1\leftarrow2$ transitions, while the highest-frequency component corresponds to the isolated $F=1\leftarrow1$ transition (see Fig.~\ref{fig6a}). 
For R(2), four hyperfine transitions ($F=2\leftarrow1$, $F=3\leftarrow2$, $F=4\leftarrow3$, $F=2\leftarrow3$) contribute to the central line, while the two side features correspond to the  $F=3\leftarrow3$ (low frequency) and $F=2\leftarrow2$ (high frequency) transition (see Fig.~\ref{fig6b}).

\begin{figure*}  
\centering
 \includegraphics[scale=0.25]{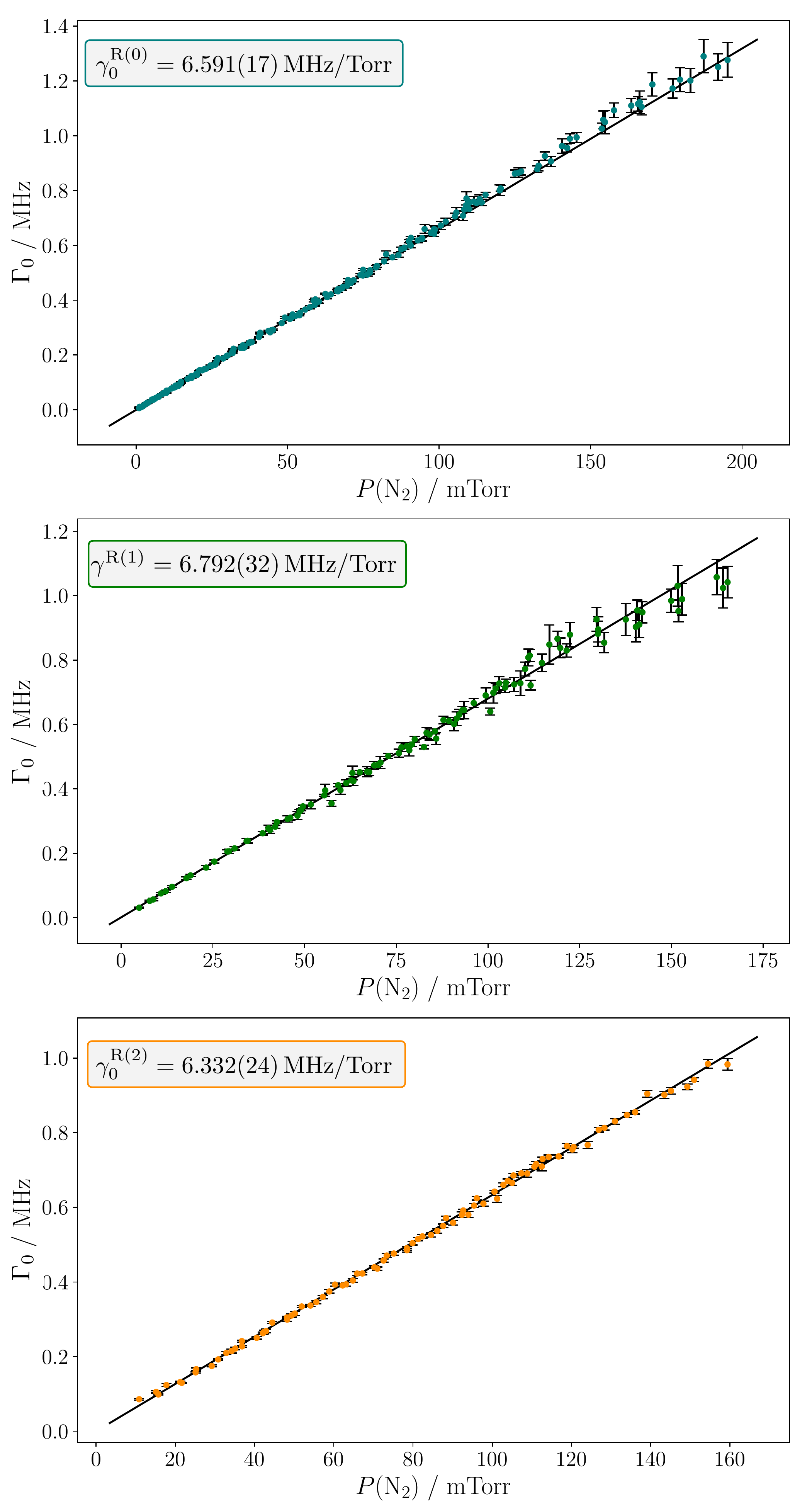}
 \caption{Linear regression analysis of $\Gamma_0$ against the pressure of \ce{N2} for the R(0), R(1) and R(2) transitions of \ce{HCN}. In all panels, three times the uncertainties retrieved from line profile analysis are shown.}
 \label{fig3} 
\end{figure*}

\begin{figure*}  
\centering
 \includegraphics[scale=0.25]{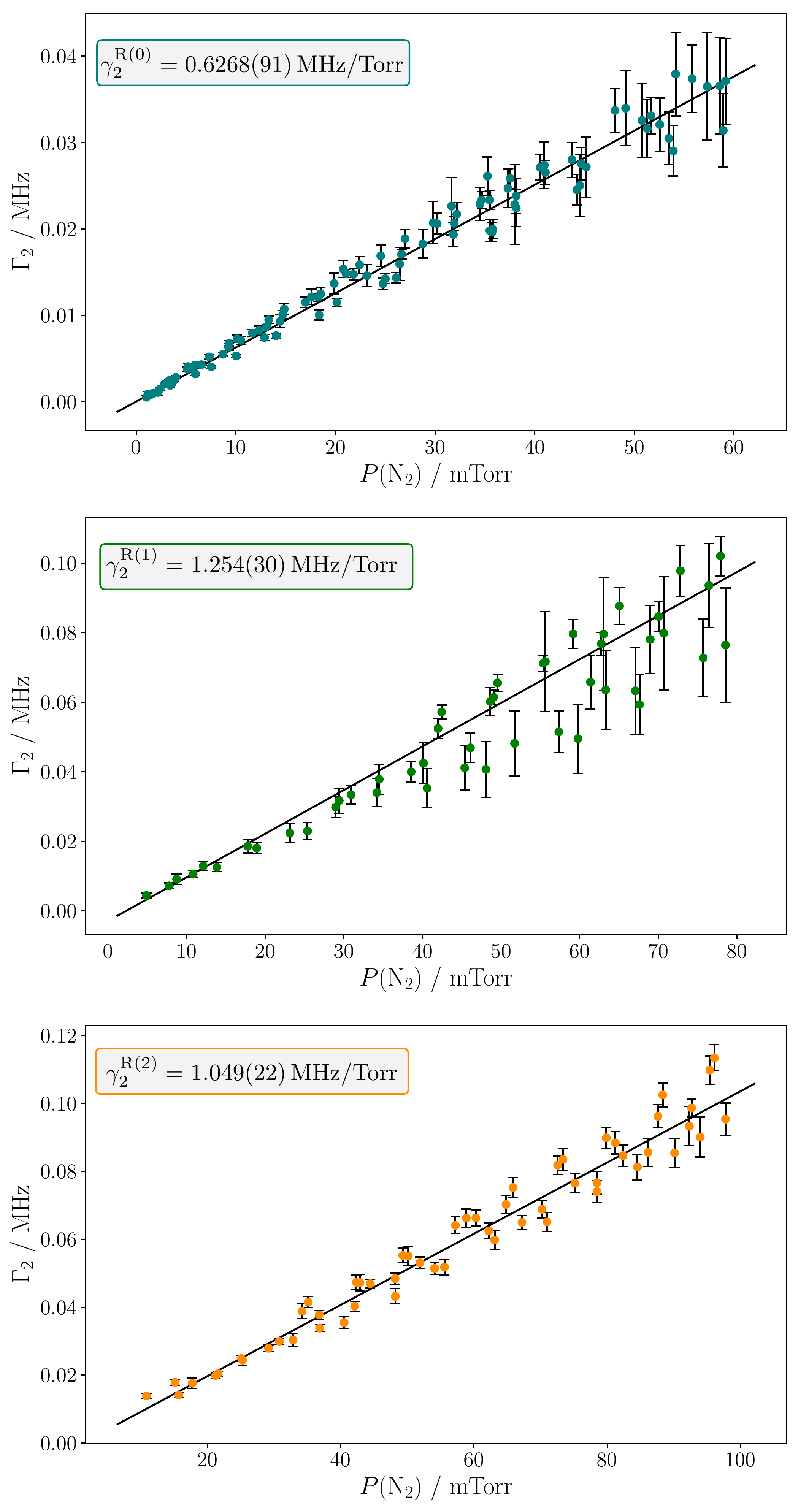}
 \caption{Linear regression analysis of $\Gamma_2$ against the pressure of \ce{N2} for the R(0), R(1) and R(2) transitions of \ce{HCN}. In all panels, three times the uncertainties retrieved from line profile analysis are shown.}
 \label{fig4} 
\end{figure*}

\begin{figure*}  
\centering
 \includegraphics[scale=0.25]{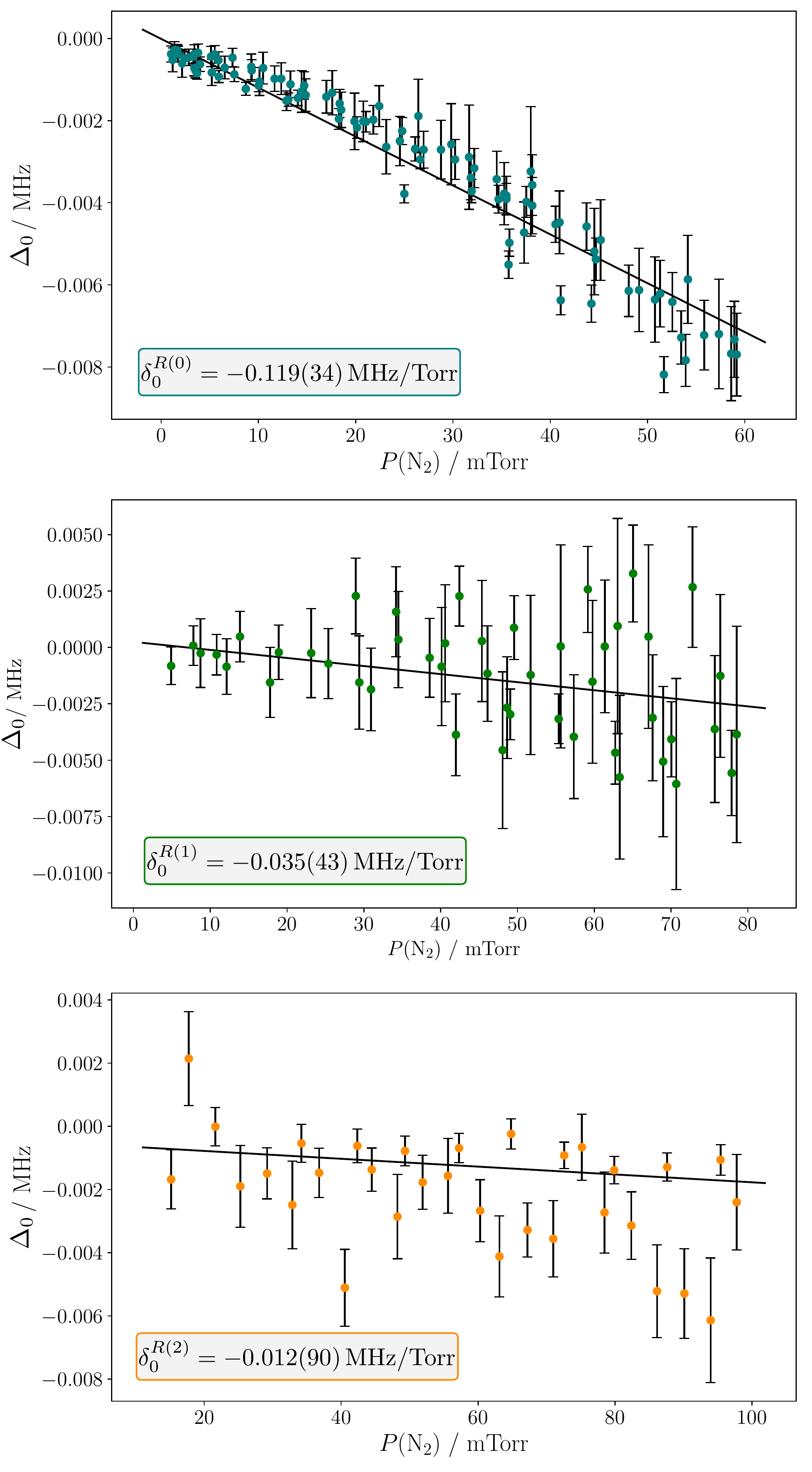}
 \caption{Linear regression analysis of $\Delta_0$ against the pressure of \ce{N2} for the R(0), R(1) and R(2) transitions of \ce{HCN}. In all panels, three times the uncertainties retrieved from line profile analysis are shown.}
 \label{fig5} 
\end{figure*}

The line profile analysis was carried out by adjusting $\Delta_0$, $\Gamma_0$ and $\phi$ for each feature of the transition. Conversely, a single $\Gamma_2$ value was used for all spectral lines to mitigate the strong correlation between fitting parameters.
Linear regression analysis of the profile parameters obtained at different \ce{N2} pressures, allowed to estimate the pressure broadening ($\gamma_0$), pressure shift ($\delta_0$) and speed-dependent broadening ($\gamma_2$) coefficients. 
For the purposes of this study, only the pressure broadening coefficient for the most intense hyperfine transition with $\Delta F=\Delta J$ was considered. 
For an absorption spectrum ($\Delta J=+1$), the dominant hyperfine transitions are $F=2\leftarrow1$ for the R(0), the $F=2\leftarrow1$, $F=3\leftarrow2$, $F=4\leftarrow3$ group for R(1), and the $F=2\leftarrow1$, $F=3\leftarrow2$, $F=4\leftarrow3$ group for R(2).
With the exception of $F=2\leftarrow1$ for R(1), these transition groups are merged into the central line, where the dominant $\Delta F=\Delta J=+1$ components are concentrated (see the relative intensities reported in the right panel of Fig.~\ref{fig_intro}). This trend becomes more and more pronounced as $J$ increases (since the hyperfine splitting is proportionally smaller), thus leading to a central line which dominates over the two $\Delta F = 0$ lateral satellites.
%
The results of the linear regression analyses for each transition are displayed in Figs.~\ref{fig3},~\ref{fig4},~\ref{fig5}.
For the R(1) and R(2) transitions, the Lorentzian half-widths determined at high \ce{N2} pressures ($P>160$\,mTorr) were excluded from the linear regression. In such conditions the increasingly broad central line partially overlaps the two weak side components, producing a unique distorted profile which complicates the line shape analysis. 
Moreover, at high pressures, the correlation between the  $\Delta_0$ and $\Gamma_2$ parameters becomes very strong and their fitted values are unstable and affected by large uncertainties. Consequently, a reduced dataset at lower pressures (up to 60\,mTorr for R(0), 80\,mTorr for R(1) and 100\,mTorr for R(2)) was used to determine these coefficients. 

\section{Theoretical details}
\label{sec3}
\subsection{Potential Energy Surface}
\label{sub3.1}
\begin{figure}[t!]
\centering
 \includegraphics[scale=0.22]{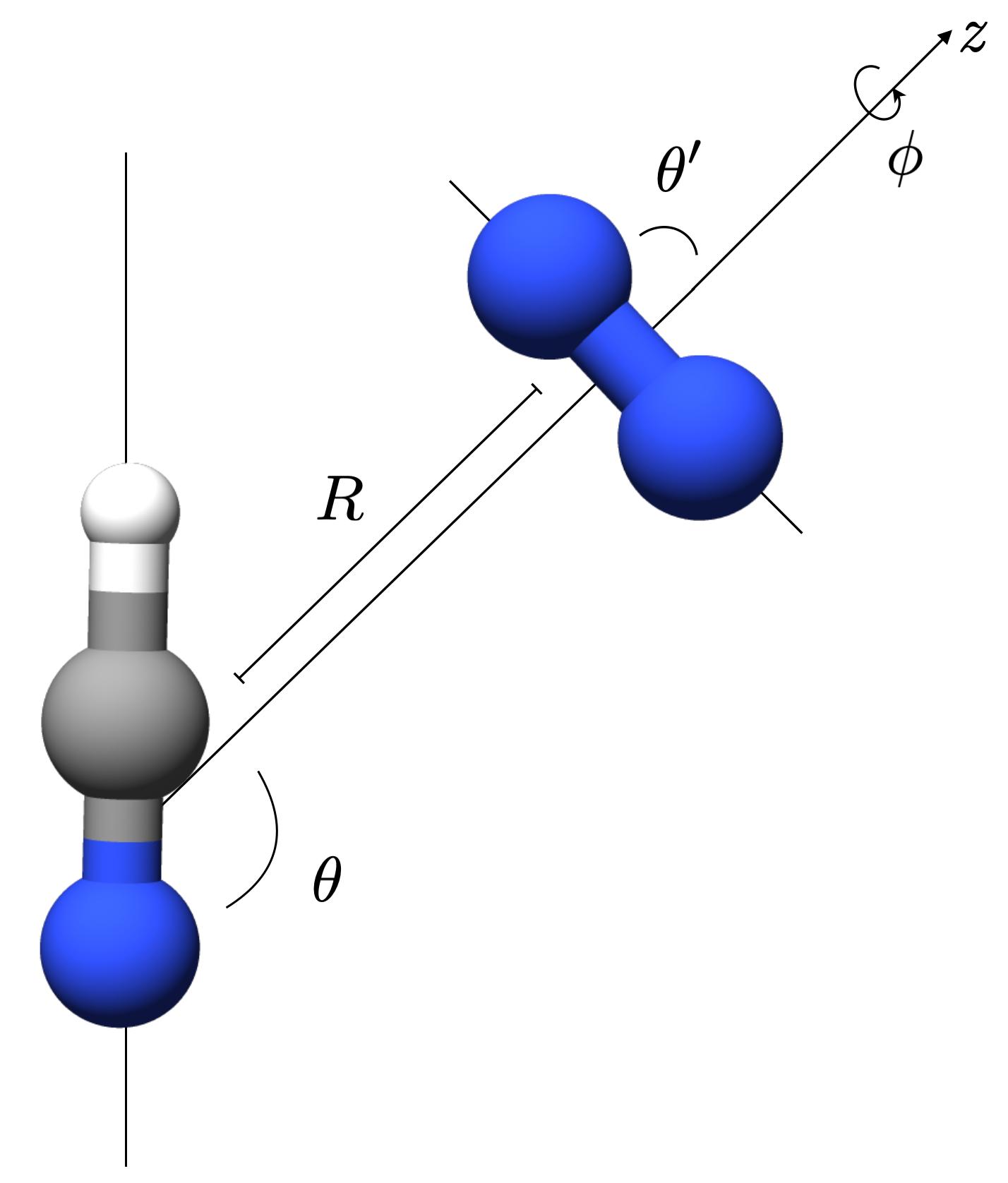}
 \caption{Jacobi internal coordinates of the \ce{HCN}$-$\,\ce{N2} collisional system. 
 }
 \label{fig1}
\end{figure}

The first step to compute the pressure induced coefficients for the \ce{HCN}--\ce{N2} system is the accurate characterization of the interaction potential between the two fragments. The potential was described by a set of four Jacobi coordinates, depicted in Fig.~\ref{fig1}, which correspond to the distance $R$ between the centers of mass of the two fragments, the $\theta$ angle between the molecular axis of HCN and the vector $\vec{R}$, the $\theta^{\prime}$ angle, which defines the orientation of \ce{N2} in the plane formed by \ce{HCN} and vector $\vec{R}$, and the $\phi$ angle, which defines the orientation of \ce{H2} out of the same plane.
Both \ce{HCN} and \ce{N2} were considered as rigid bodies fixed at their equilibrium geometries, with $r(\ce{CH})=1.0606$\,\AA, $r(\ce{CN})=1.1533$\,\AA\,~\citep{piccardo2015semi} and $r(\ce{NN})=1.09768$\,\AA\,~\citep{huber1979molecular}.
The electronic interaction energies were subsequently computed over a grid of Jacobi coordinates, purposely chosen to accurately sample the anisotropy of the system. In detail, 3345 points have been computed for $R$ ranging between 2 an 12\,\AA\, for 25 equally spaced values of $\theta$, and for 5 orientations of \{$\theta^{\prime},\phi$\} coordinates. 
The reason why only five orientations of \ce{N2} with respect to the target molecule were considered will be detailed afterwards.
All the calculations employed explicitly correlated coupled-cluster theory~\citep{adler2007simple,knizia2009simplified,peterson2008systematically} including singles, doubles, and a perturbative treatment of triple excitations~\citep{raghavachari1989fifth}, i.e., the CCSD(T)-F12a method, in combination with the aug-cc-pVQZ basis set~\citep{dunning2001gaussian, woon1993gaussian}. This level of theory has already been proven to yield very good performances in computing the electronic energies of interacting fragments, still maintaining an affordable computational cost (as an example, the reader is referred to the benchmark studies of~\citet{ajili2013accuracy} and~\citet{tonolo2021improved}).
For all the calculations, the MOLPRO suite of programs\footnote{\url{https://www.molpro.net}.}~\citep{werner2012wires} was employed.

For each set of Jacobi coordinates, the interaction energies ($E_\text{int}$) corrected for the basis set superposition error (BSSE) by means of the counterpoise (CP) correction~\citep{boys1970calculation} were determined as follows:
\begin{equation}
\begin{aligned}
E_\text{int} &= E_\text{AB} - (E_\text{A} + E_\text{B}) + \Delta E_{\text{CP}}\,, \\ 
\text{where} \qquad \Delta E_{\text{CP}}&= (E^{\text{AB}}_{\text{A}} - E^{\text{A}}_{\text{A}}) + (E^{\text{AB}}_{\text{B}} - E^{\text{B}}_{\text{B}})\,.
\end{aligned}
\end{equation}
Here, $E^{\text{AB}}_{\text{X}}$ is the energy of the monomer calculated with the basis set used for the cluster and $E^{\text{X}}_{\text{X}}$ is the energy of the monomer computed with its own basis set ($\text{X} = \text{A},\text{B}$).

Having computed the \emph{ab initio} points, in order to solve the nuclear Schrödinger equation which describes the quantum scattering problem, the interaction potential needs to be fitted as an expansion over angular functions. For an interaction between two linear rigid rotors, such expansion can be written as~\citep{green1975rotational,wernli2007rotational,wernli2007rotationalb}:
\begin{equation}\label{fit}
V\left(R, \theta, \theta^{\prime}, \phi\right)=\sum_{l_1, l_2, \mu \geq 0} v_{l_1 l_2 \mu}(R) P_{l_1}(\theta)\left[Y_{l_2-\mu}\left(\theta^{\prime}, \phi\right)+(-1)^\mu Y_{l_2 \mu}\left(\theta^{\prime}, \phi\right)\right]\,.    
\end{equation}
Here, $v_{l_1 l_2 \mu}(R)$, $P_{l_1}(\theta)$, and $Y_{l_i \mu}(\theta^\prime,\phi)$ are sets of radial coefficients, Legendre polynomials, and spherical harmonics, respectively. They are associated to $l_1$, $l_2$ and $\mu$, which are indices linked with the rotational angular moments of \ce{HCN} ($j_1$), \ce{N2} ($j_2$) and of the system ($j_{12}$), respectively (see~\citet{green1975rotational,brown2003rotational,edmonds2016angular} for further details).\footnote{Unlike the commonly used spectroscopic notation adopted in the rest of the manuscript, we use in this Section lower-case letter $j$ to label the rotational quantum number of the isolated molecules, while for the entire collisional system, the upper-case letter $J$ is employed. In details, $j_1$ and $j_2$ correspond to the angular moments of \ce{HCN} and \ce{N2}, respectively, and $j_a$ and $j_b$ refer to the initial and final states of the $J^\prime\leftarrow J$ rotational transition.} 

The $l_2$ index reflects the anisotropy of the potential with respect to the orientation of the perturber. 
For the target system, only the terms with $l_2 \leq 2$ were considered in the basis expansion.
This approximation was already found to be appropriate for similar systems with homonuclear perturbers~\citep{faure2005role,wernli2006improved,tonolo2024collisional} and allows for a simplified representation of the analytical interaction potential.
Accordingly to this approximation, the only basis functions that describe the dependence of the potential on the relative orientations of \ce{N2} are the $Y_{00}$, $Y_{20}$, $Y_{21}$ and $Y_{22}$ spherical harmonics. 
Hence, at fixed $R$ and $\theta$ values, the dependence of the potential on the orientation of the perturber can be entirely described from the knowledge of only four orientations $\{\theta',\phi\}$. 
By choosing five sets of $\{\theta',\phi\}$ orientations we described the dependence of the potential on them and provided an over-determined system to test the accuracy of the $l_2$ truncation (see, for details,~\citet{wernli2006improved, tonolo2022hyperfine})\@.
\begin{figure*}  
 \includegraphics[scale=0.42]{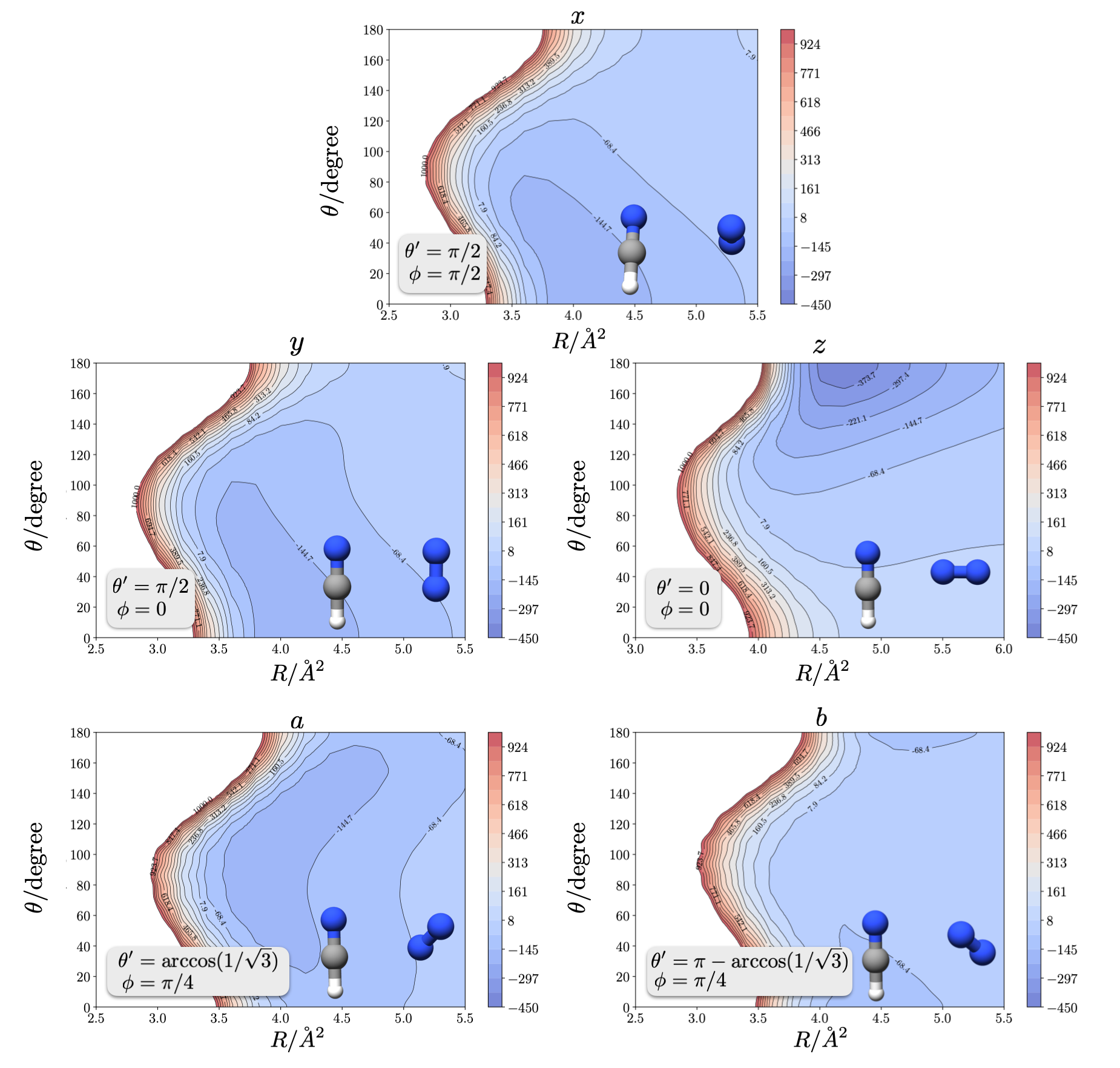}
 \caption{Contour plots of the \ce{HCN}--\ce{N2} interaction PES for five different orientations of \ce{N2}.}
 \label{fig2} 
\end{figure*}
The chosen orientations were those selected by~\citet{wernli2007rotational} for the \ce{HC3N}-\ce{H2} system:
\begin{subequations}
  \begin{align} 
  x & \rightarrow\left(\theta^{\prime}=\frac{\pi}{2}, \phi=0\right)\,; \\
  y & \rightarrow\left(\theta^{\prime}=\frac{\pi}{2}, \phi=\frac{\pi}{2}\right)\,; \\
  z & \rightarrow(\theta^{\prime}=0, \phi=0)\,; \\
  a & \rightarrow\left(\theta^{\prime}=\arccos \left(\frac{1}{\sqrt{3}}\right), \phi=\frac{\pi}{4}\right)\,; \\
  b & \rightarrow\left(\theta^{\prime}=\pi-\arccos \left(\frac{1}{\sqrt{3}}\right), \phi=\frac{\pi}{4}\right)\,.
  \end{align}
 \label{xyzab}%
\end{subequations}
Each orientation of Eq.~\eqref{xyzab} was separately fitted as an expansion over $P_{\lambda}$ Legendre polynomials within the following expression~\citep{lique2019gas}:
\begin{equation} \label{pot}
 V\left(R, \theta \right)=\sum_{\lambda}  v_{\lambda} (R) P_{\lambda} \left(\cos \theta \right) \,.
\end{equation}
The $v_{\lambda} (R)$ radial coefficients were fitted to a functional form which takes into account the sizable contribution due to induction interactions of the \ce{HCN-N2} system~\citep{tang1984improved}:
\begin{multline}  
\label{vexp}
 v_\lambda(R) = \text{e}^{-a_1^\lambda R}\left(a_2^\lambda + a_3^\lambda R + a_4^\lambda R^2 + a_5^\lambda R^3\right) \\
 -\frac{1}{2}\left[1 + \tanh\left(R/R_\text{ref}\right)\right]
 \left(\frac{C^{\lambda}_4}{R^4} + \frac{C^{\lambda}_6}{R^6} + \frac{C^{\lambda}_8}{R^8} + \frac{C^{\lambda}_{10}}{R^{10}}\right) \,,
\end{multline}
where $a^{\lambda}_n$ denotes the coefficients of the short-range region ($0 < R < R_\text{ref}$) and $C^{\lambda}_n$ the $R^{-n}$ terms in the long-range extrapolated domain ($R > R_\text{ref}$). For each angular dependency block, all coefficients and the $R_\text{ref}$ value were optimized in the fit. 

For each orientation, the fitted points resulted in good agreement with the corresponding \textit{ab initio} computed ones, with deviations on average within 2\% over the entire grid. 
The energy plots corresponding to the chosen \ce{N2} orientations are shown in Fig.~\ref{fig2}. 
The weak anisotropy of the potential with respect to the $\{\theta^{\prime},\phi\}$ coordinates validated the choice of truncating the potential to the $l_2\leq 2$ terms.
Hence, the global four dimension (4D) potential was introduced $via$ the following equation~\citep{wernli2006improved}:
\begin{equation}
\begin{aligned}
V(R, \theta, \theta^{\prime},\phi)&= V_{\text{av}}(R, \theta)+\\&+\frac{1}{2}\left[V(R,\theta,z)-V_{\text{av}}(R, \theta)\right]\left(3 \cos ^{2} \theta^{\prime}-1\right)+\\&+\frac{3}{2}\left[V(R,\theta,a)-V(R,\theta,b)\right] \sin \theta^{\prime} \cos \theta^{\prime} \cos \phi+\\&+\frac{1}{2}\left[V(R,\theta,x)-V(R,\theta,y)\right] \sin ^{2} \theta^{\prime} \cos \left(2 \phi\right)\,,\\
\text{where}\, \qquad
V_{\text{av}}(R, \theta)&= \frac{1}{7} [ 2\left(V(R, \theta, a)+V(R, \theta, b)\right)+\\ &+ \left(V(R, \theta, x)+V(R, \theta, y)+V(R, \theta, z)\right)]\,. 
\end{aligned}
\label{angintr}
\end{equation}

The 4D interaction potential of \ce{HCN} with \ce{N2} exhibits a global minimum when the nitrogen molecule is aligned with \ce{HCN} on the hydrogen side, i.e., for the $\{\theta=180;\theta^\prime=0;\phi=0$\} set of coordinates. Indeed, the resulting van der Waals interaction greatly stabilizes the complex, whose energy at $R=4.7$\,\AA\, is $-432.33$\,cm$^{-1}$. For the other orientations, the interaction is weaker, reaching maximum values of about $E=-204$\,cm$^{-1}$ when the nitrogen is oriented perpendicular to the \ce{HCN} molecular axis and on the side of the nitrogen atom. 

\subsection{\textit{Ab initio} calculations of the spectral line shape parameters}
\label{sub3.2}

The calculation of pressure broadening coefficients for the N$_{2}$-perturbed HCN lines relies on determining the scattering $S$-matrices by solving the close-coupling equations using the HCN-N$_{2}$ PES. Both molecules were treated as rigid rotors. While the quantum scattering theory for two rigid rotors is well established~\citep{green1975rotational, Alexander_1977}, in the following, we provide a summary of the key aspects of our quantum scattering calculations and introduce the new approximate approach employed in this study to enhance computational efficiency.

Although the hyperfine splitting is shaping the experimental spectrum (see Figs.~\ref{fig6},~\ref{fig6a} and~\ref{fig6b}), a full study of hyperfine-resolved line shape parameters is beyond the scope of this work. In principle, such calculations can be performed using the recoupling technique, as demonstrated for mo\-le\-cule–\-a\-tom systems~\citep{Green1988,Buffa2011a,Buffa2011b}, and it can be shown that, except for the R(0) transition, the line shape parameters would vary for each hyperfine component~\citep{Green1988}. Since the experiments did not resolve all hyperfine components (with parameters only for the strongest components or for blended peaks being reported), we provided theoretical line shape parameters for hyperfine-free rotational $j_{b}\leftarrow j_{a}$ transitions.

Following previous work on linear molecules colliding with N$_{2}$~\citep{gancewski2021} or O$_{2}$~\citep{zadrozny2022, Olejnik_2023}, we solved the quantum scattering equations in the body-fixed frame of reference, leveraging the predominantly block-diagonal structure of the coupling matrix~\citep{Launay_1977}. This choice significantly reduces computational time and memory requirements. The close-coupling equations were integrated numerically using the log-derivative propagator implemented in the \texttt{BIGOS} scattering code, developed by the Toruń group~\citep{Jozwiak_2024, Jozwiak_2024_code}. At sufficiently large intermolecular separations, the log-derivative matrix was transformed into the space-fixed frame, and the scattering $S$-matrix was obtained by applying standard boundary conditions to the asymptotic form of the scattering wavefunction.

The scattering $S$-matrices were then used to determine the complex, generalized spectroscopic cross-sections, which describe the collisional perturbation of the $j_{b} \leftarrow j_{a}$ transition of HCN by the N$_{2}$ molecule in the $j_{2}$ rotational state:
 \begin{align}
 \begin{split}
   &\sigma_{\lambda}^{\kappa} (j_{a}, j_{b}, j_{2}, E_{\mathrm{kin}}) = \frac{\pi}{k^2} \sum_{j_{2}'} \sum_{L, L', \bar{L}, \bar{L}'} \sum_{\mathcal{J}_{a}, j_{a2}, j_{a2}'} \sum_{\mathcal{J}_{b}, {j}_{b2}, {j}_{b2}'} i^{L-L'-\bar{L}+\bar{L}'}
   \\ &\times
     (-1)^{\lambda + j_2 - j_2' + L - L' - \bar{L} + \bar{L}'}
     [\mathcal{J}_{a}, \mathcal{J}_{b}] \sqrt{[L,L',\bar{L},\bar{L}',j_{a2},j_{a2}',{j}_{b2},{j}_{b2}']}
     \\
     &\times 
     \begin{pmatrix}
        L       & \bar{L}  & \lambda \\
        0       & 0        & 0
     \end{pmatrix} \begin{pmatrix}
        L'      & \bar{L}' & \lambda \\
        0       & 0        & 0
     \end{pmatrix}
     \begin{Bmatrix}
        \kappa       & j_{a2}'       & {j}_{b2}' \\
        j_2'    & j_f     & j_i
     \end{Bmatrix} \begin{Bmatrix}
        \kappa       & {j}_{b2}  & j_{a2} \\
        j_2     & j_a      & j_b
     \end{Bmatrix}
     \\ &\times \begin{bmatrix}
        j_{a2}       & j_{a2}'       & \bar{L}   & \bar{L}' \\
        {j}_{b2} & L        & {j}_{b2}'  & L'      \\
        \kappa       & \mathcal{J}_{b}      & \mathcal{J}_{a}       & \lambda
     \end{bmatrix} 
     \bigg(
        \delta_{j_2 j_{2}'} \delta_{j_{a2} j_{a2}'} \delta_{{j}_{b2} j_{b2}'} \delta_{L L'} \delta_{\bar{L} \bar{L}'} \\
        &\,\,\,\,\,\,\,\,- S^{\mathcal{J}_{a}}(E_{T_{a}})_{(j_a j_2') j_{a2}' L' ; (j_a j_2) j_{a2} L }
        S^{\mathcal{J}_{b}\,*}(E_{T_{b}})_{(j_b j_2') j_{b2}' \bar{L}' ; (j_b j_2) j_{b2} \bar{L} }
     \bigg). \label{eq:crosssection}
 \end{split}
 \end{align}
 The parameter $\lambda$ distinguishes between the two types of cross-sections computed in this work. When $\lambda=0$, the real and imaginary parts of this cross-section describe pressure broadening and pressure shift of the spectral line, respectively~\citep{Ben_Reuven_1966a, Ben_Reuven_1966b}. Accordingly, we refer to $\mathrm{Re}(\sigma^{\kappa}_{\lambda=0})$ as the pressure broadening cross-section and $\mathrm{Im}(\sigma^{\kappa}_{\lambda=0})$ as the pressure shift cross-sections. For $\lambda=1$, the complex cross-section describes the collisional perturbation of the translational motion of the molecule undergoing a spectral transition~\citep{Corey_1984, monchick1986diatomic}. This effect plays a crucial role in describing beyond-Voigt effects, such as Dicke narrowing~\citep{dicke1953}. Following previous works, we refer to $\sigma^{\kappa}_{\lambda=1}$ as the Dicke cross-section. The parameter $\kappa$ represents the tensorial rank of the radiation-matter interaction. For the electric dipole transitions considered here, $\kappa=1$. In Eq.~\eqref{eq:crosssection}, pre- and post-collisional quantum numbers are denoted without and with primes, respectively. The formula involves the coupling of eight angular momenta in total: the two rotational angular momenta specifying the rotational transition, $\vec{\boldsymbol{j}}_{a}$ and $\vec{\boldsymbol{j}}_{b}$, the pre- and post-collisional rotational angular momenta of the perturbing molecule, $\vec{\boldsymbol{j}}_{2}$ and $\vec{\boldsymbol{j}}_{2}'$, and four angular momenta describing relative orbital motion of the two molecules: $\vec{\boldsymbol{L}}$, $\vec{\boldsymbol{L}}'$, $\vec{\bar{\boldsymbol{L}}}$ and $\vec{\bar{\boldsymbol{L}}}'$. The rotational angular momenta of the two molecules are coupled to form the total rotational angular momenta of the pair:
 \begin{equation}
\vec{\boldsymbol{j}}_{a2}=\vec{\boldsymbol{j}}_{a}+\vec{\boldsymbol{j}}_{2},\,\,\,     \vec{\boldsymbol{j}}_{a2}'=\vec{\boldsymbol{j}}_{a}+\vec{\boldsymbol{j}}_{2}',\,\,\,\vec{\boldsymbol{j}}_{b2}=\vec{\boldsymbol{j}}_{b}+\vec{\boldsymbol{j}}_{2},\,\,\,\vec{\boldsymbol{j}}_{b2}'=\vec{\boldsymbol{j}}_{b}+\vec{\boldsymbol{j}}_{2}'.
 \end{equation} 
 The total rotational angular momenta are coupled with the respective relative orbital angular momenta, to form the total angular momenta of the system:
 \begin{equation}
     \vec{\boldsymbol{\mathcal{J}}}_{a} = \vec{\boldsymbol{j}}_{a2} +  \vec{\boldsymbol{L}},\,\,\,\vec{\boldsymbol{\mathcal{J}}}_{a} = \vec{\boldsymbol{j}}_{a2}' +  \vec{\boldsymbol{L}}', \,\,\,\vec{\boldsymbol{\mathcal{J}}}_{b} = \vec{\boldsymbol{j}}_{b2} +  \vec{\bar{\boldsymbol{L}}},\,\,\,\vec{\boldsymbol{\mathcal{J}}}_{b} = \vec{\boldsymbol{j}}_{b2}' +  \vec{\bar{\boldsymbol{L}}}' .
 \end{equation}
 The shorthand notation $[x_{1}, ... x_{n}] = (2x_{1} + 1)\cdot ... \cdot(2x_{n}+1)$ represents multiplicities, while quantities  $(\ldots)$, $\{\ldots\}$ and $[\ldots]$ are Wigner 3-$j$, 6-$j$ and 12-$j$ symbols, respectively~\citep{Yutsis}.  Note that Eq.~\eqref{eq:crosssection} implies that two distinct scattering $S$-matrices are needed to obtain $\sigma^{\kappa}_{\lambda}$ for a given collision energy, $E_{\rm{kin}}$. The first matrix is computed at the total energy ${E_{T_{a}} = E_{\rm{kin}}+E_{j_{a}}+E_{j_{2}}}$, and the second matrix is calculated at ${E_{T_{b}} = E_{\rm{kin}}+E_{j_{b}}+E_{j_{2}}}$.

The complex cross-sections, $\sigma^{\kappa}_{\lambda=0}$, once averaged over the Max\-well\--\-Boltz\-mann distribution of collision energies, and the Boltzmann population distribution of the perturbing molecule provide the pressure broadening, $\gamma_{0}$, and shift, $\delta_{0}$, coefficients:
\begin{equation}
\label{eq:gamma0}
    \gamma_{0} - i\delta_{0} = \frac{1}{2\pi c} \frac{\langle v_{r}\rangle}{k_{\mathrm{B}}T} \sum_{j_{2}} p_{j_{2}} \int_{0}^{+\infty} x \exp(-x) \sigma^{\kappa}_{\lambda=0}(j_{a},j_{b}, j_{2};E_{\mathrm{kin}}=xk_{\mathrm{B}}T) \mathrm{d}x .
\end{equation}
For the reasons explained in Section~\ref{sec:open_channels}, the pressure shift coefficient was not determined in the present work. In Eq.~\eqref{eq:gamma0}, $c$ is the speed of light in vacuum, $k_{\mathrm{B}}$ is the Boltzmann constant, and $T$ is the temperature. The mean relative speed of the colliding pair is given by $\langle v_{r}\rangle = \sqrt{8k_{\mathrm{B}}T/(\pi \mu)}$, where $\mu$ is the reduced mass of the system. The fractional population of the perturbing molecule in a given $j_{2}$ rotational state is denoted by $p_{j_{2}}$ and is given by:
\begin{equation}
    p_{j_2}(T) = \frac{1}{Z(T)} w_{j_{2}} (2j_2+1) \exp(-E_{j_2}/(k_BT)) \text{,} \label{eq:pN2}
\end{equation}
where $E_{j_2}$ is the energy of the considered rotational state, and
\begin{equation}
Z(T) = \sum_{j_2}w_{j_{2}}(2j_2+1) \exp(-E_{j_2}/(k_BT)) \label{eq:ZT}
\end{equation}
is the partition function. The weighting factor $w_{j_{2}}$ accounts for the nuclear spin statistics. As a homonuclear molecule, nitrogen exists in two spin isomeric forms, distinguished by the total nuclear spin, $I_{T}$, which results from coupling the two spins of the $^{14}$N nuclei ($I=1$): \textit{ortho}-N$_{2}$ ($I_{T}=0, 2$) and \textit{para}-N$_{2}$ ($I_{T}=1$). The \textit{ortho}-N$_{2}$ isomer comprises even $j_{2}$ values, while \textit{para}-N$_{2}$ includes rotational states with odd $j_{2}$. Consequently, $w_{j_{2}}=6$ for even $j_{2}$, and $w_{j_{2}}=3$ for odd $j_{2}$. The two \textit{ortho} and \textit{para} isomers of \ce{N2} are not collisionally coupled and can therefore be treated independently in scattering calculations. 

The HCN spectra exhibit speed-dependence of collisional broadening (see Sec.~\ref{sub2.2}). In the experimental analysis, a quadratic dependence on speed was assumed, which is characterized by the $\gamma_{2}$ parameter. This parameter can also be evaluated \textit{ab initio} from the complex cross-sections, $\sigma^{q}_{\lambda=0}$, averaged over the appropriate conditional probability distribution (see Ref.~\citep{Stankiewicz_2021} for details):
\begin{align}
\begin{split}
\label{eq:gamma2}
\gamma_2=\frac{1}{2\pi c}\frac{1}{k_B T}&\frac{\left<v_r\right>\sqrt{M_a}}{2}\exp(-y^2) \sum_{j_{2}} p_{j_{2}} \\
\times& \int\limits_{0}^{\infty}\left(2\bar{x}\cosh(2\bar{x}y)-\left(\frac{1}{y}+2 y\right)\sinh(2\bar{x}y)\right)\\
\times&\, \bar{x}^2\exp(-\bar{x}^2)\mathrm{Re}\Bigl(\sigma^{\kappa}_{\lambda=0}(j_{a},j_{b}, j_{2};E_{\mathrm{kin}}=\mu (\bar{x}\bar{v}_{p})^{2}/2)\Bigr)d\bar{x},
\end{split}
\end{align}
where $v_r$ is the relative (absorber with reference to perturber) speed of the colliding mo\-le\-cules, $\langle v_{r} \rangle$ is its mean value at temperature $T$, $\bar{v}_{p}$ is the most probable speed of the perturber, $M_{a} = \frac{m_{a}}{m_{a}+m_{p}}$, $\bar{x}=\frac{2v_r}{\sqrt{\pi M_a}\left<v_r\right>}$, $y=\sqrt{\frac{m_p}{m_a}}$ and $m_{a}$ and $m_{p}$ are masses of the active molecule, and the perturber, respectively.

Finally, although the spectra reported in this work do not exhibit Dicke narrowing, we provide, for the first time, reference values for the complex Dicke parameter, defined as:
\begin{align}
\begin{split}
        \tilde{\nu}_{opt}^r -  i \tilde{\nu}_{opt}^i = \frac{1}{2\pi c} \frac{\langle v_r \rangle M_a}{k_B T}  \sum_{j_{2}} p_{j_{2}}&
        \int \limits_{0}^{\infty} x \exp(-x) \Bigl(\frac{2}{3}x\sigma_{\lambda=1}^{\kappa}(j_{a},j_{b}, j_{2};E_{\mathrm{kin}}=xk_{\mathrm{B}}T)  \\ -&\sigma_{\lambda=0}^{\kappa}(j_{a},j_{b}, j_{2};E_{\mathrm{kin}}=xk_{\mathrm{B}}T)\Bigr) dx . 
\label{eq:nuopt}
\end{split}
\end{align}
For the same reasons why we did not provide theoretical values for the pressure shift coefficient and its speed-dependence, we computed only the real part of the Dicke parameter. Test calculations indicate that the real part is at least an order of magnitude larger than the imaginary part.

For practical reasons, both the integration range over collision energies and the summation over the population distribution of  $j_{2}$ in Eqs.~\eqref{eq:gamma0},~\eqref{eq:gamma2} and~\eqref{eq:nuopt} must be truncated at certain upper limits, $E_{\mathrm{kin}}^{\mathrm{max}}$ and $j_{2}^{\mathrm{max}}$. Ideally, these limits should be set high enough to capture over 99\% of the respective distributions. However, achieving this level of coverage exceeds the feasibility of fully converged quantum scattering calculations (see Section~\ref{sec:open_channels}). 

\begin{figure}[!ht]
    \centering
    \includegraphics[width=\linewidth]{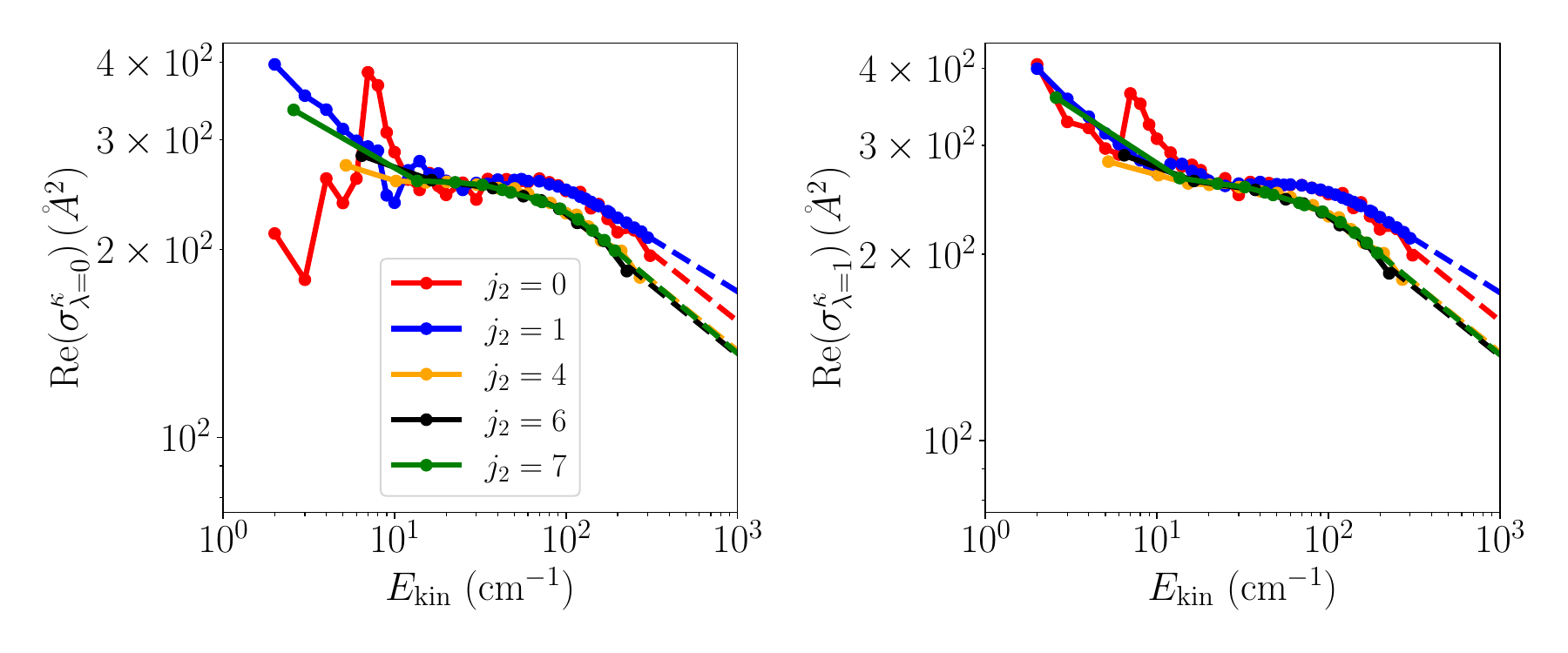}
    \caption{Pressure broadening cross-sections (left panel) and the real part of the Dicke cross-sections (right panel) for the N$_{2}$-perturbed R(1) transition of HCN calculated for all $j_{2} \leq 7$, with a subset of these results displayed for clarity. Dots represent \textit{ab initio} values, while the dashed lines are the extrapolations (see the text for details).}
    \label{fig:XS}
\end{figure}

To address this, we adopted an extrapolation approach based on the smooth behavior of the generalized spectroscopic cross-sections at large collision energies and the apparent lack of significant variation in cross-sections for large values of $j_{2}$, as demonstrated in previous studies~\citep{Jozwiak_2021, gancewski2021, zadrozny2022, Olejnik_2023}. First, we computed cross-sections for $j_{2}$ values up to $j_{2}=7$. For clarity, only a subset of these results is shown in Fig.~\ref{fig:XS}, where we present the pressure broadening and the real part of the Dicke cross-sections for the N$_{2}$-perturbed R(1) line in HCN for selected values of $j_{2}$. We observed that, at sufficiently high collision energies, the cross-sections follow a power-law dependence:
\begin{align}
    \text{Re} \left[ \sigma^{\kappa}_{\lambda}(E_{\mathrm{kin}}) \right] =  \frac{A}{E_{\mathrm{kin}}^b}   .\label{eq:powerlaw}
\end{align}
The parameters $A$ and $b$ were determined by fitting this functional form to the last three data points in the computed energy range, where the cross-sections exhibit a smooth power-law behavior. Thus, in the actual evaluation of Eqs.~\eqref{eq:gamma0},~\eqref{eq:gamma2}, and~\eqref{eq:nuopt}, we combined \emph{ab initio} values with extrapolated cross-sections using this power-law behavior, ensuring that over 99\% of the collision energy distribution was properly accounted for.

Additionally, Fig.~\ref{fig:XS} demonstrates that, in the high-kinetic energy regime, the cross-sections for $j_{2}\geq 4$ exhibit differences smaller than 1\% compared to the adjacent $j_{2}$ values. Based on this observation, when evaluating the line shape parameters, we used the computed cross-sections up to $j_{2}=7$ and assumed that, for $j_{2} \geq 7$, the cross-sections can be approximated as $\sigma_{\lambda}^{\kappa}(j_{a},j_{b}, j_{2}=7;E_{\mathrm{kin}}$). To ensure that the sum over $j_{2}$ covers more than 99\% of the perturber's population distribution, the cross-sections were extrapolated up to $j_{2}=22$.

\subsubsection{The open channels approach}
\label{sec:open_channels}
The accuracy of the computed line shape parameters depends, among other factors, on the number of scattering channels included in quantum scattering calculations. To simplify the discussion, we adopt the space-fixed description of the scattering problem. A single scattering channel corresponds to a specific internal state of the two colliding molecules ($j_{1}$ and $j_{2}$) and a particular value of the relative orbital angular
momentum ($L$).  For rigid rotor molecules, a channel is defined as:
\begin{equation}
    |\xi^{\mathcal{J}}\rangle = |((j_{1}j_{2})j_{12}L)\mathcal{J}\rangle.
\end{equation}
Here, the subscript indicates that the quantum scattering problem is diagonal with respect to the total angular momentum, $\mathcal{J}$.\footnote{Recall that the quantum scattering problem is additionally block-diagonal with respect to spatial parity. This means that for a single $\mathcal{J}$ block one can solve the close-coupling equations for two parity blocks indepedently.} Note that we do not explicitly write its projection on the quantization axis, $M$, as the close-coupling equations do not conserve this quantum number.

A key quantity that distinguishes two types of scattering channels, is the asymptotic energy associated with each channel, defined as:
\begin{equation}
    E_{\xi}^{\mathrm{asympt}} = E_{j_{1}} + E_{j_{2}},
\end{equation}
which represents the energy at which the effective interaction potential tends to as $R\rightarrow \infty$. When two diatomic molecules are initially (before the collision) in states $j_{1_{i}}$ and $j_{2_{i}}$, and they collide with kinetic energy $E_{\mathrm{kin}}$, the scattering channels can be classified as either asymptotically open,
    \begin{equation}
        E_{j_{1_{i}}}+E_{j_{2_{i}}}+E_{\mathrm{kin}} \geq E_{\xi}^{\mathrm{asympt}},
    \end{equation}
or asymptotically closed,
    \begin{equation}
        E_{j_{1_{i}}}+E_{j_{2_{i}}}+E_{\mathrm{kin}} <E_{\xi}^{\mathrm{asympt}} .
    \end{equation}
It is important to note that channels with different $j_{12}$, $L$ and $\mathcal{J}$, but the same $j_{1}$ and $j_{2}$, are degenerate at $R\rightarrow\infty$.

To ensure convergence of cross-sections---whether standard state-to-state or generalized cross-sections---all open channels must be included in the calculations, along with a carefully chosen set of closed channels. Selecting the appropriate closed channels is a non-trivial task, as their relevance depends on several factors, including collision energy. In general, the contribution of a closed channel increases as the energy gap $E_{\xi}^{\mathrm{asympt}}-(E_{j_{1_{i}}}+E_{j_{2_{i}}}+E_{\mathrm{kin}} )$ decreases--- i.e., channels that are only slightly closed are more likely to contribute significantly to the magnitude of the cross-sections. Additionally, the importance of a given closed channel depends on the strength of the anisotropic coupling of the initial state ($j_{1_{i}}$, $j_{2_{i}}$) with that closed channel. If the coupling is determined solely by the interaction potential, the relevance of a particular closed channel associated with internal states ($j_{1}, j_{2}$) can be assessed by examining the anisotropic terms in the PES expansion, Eq.~\eqref{fit}, $v_{l_1 l_2 \mu}(R)$, while considering the selection rules:
\begin{equation*}
    |j_{1_{i}}-j_{1}| \leq l_{1} \leq j_{1_{i}} + j_{1},\,\,\,
    |j_{2_{i}}-j_{2}| \leq l_{2} \leq j_{2_{i}} + j_{2} .
\end{equation*}

To assess the convergence of the pressure broadening coefficients with respect to the number of channels (or, equivalently, to the size of the rotational basis) we performed a series of quantum scattering calculations at a selected collision energy of $E_{\mathrm{kin}} = 100$~cm$^{-1}$. These calculations were carried out for selected total angular momentum blocks $(\mathcal{J}_{a}=0, \mathcal{J}_{b}=0,1$) to obtain the generalized spectroscopic cross-sections, $\sigma^{\kappa}_{\lambda}$, for the N$_{2}$-perturbed R(0) transition of HCN, assuming the perturber is initially in its rotational ground state  ($j_{2}=0$).

Each test was performed with a different number of channels, controlled by the largest rotational quantum numbers of HCN and N$_{2}$ included in the calculations, denoted as $j_{1}^{\mathrm{max}}$ and $j_{2}^{\mathrm{max}}$, respectively. The results indicated that achieving a precision of a few percent in the computed pressure broadening cross-section and the real part of the Dicke cross-section requires including rotational levels up to $j_{1}^{\mathrm{max}}=15$ and $j_{2}^{\mathrm{max}}=14$. This corresponds to approximately 15~000 scattering channels per each total angular momentum and parity block, which is at the very limit of what can be rigorously computed, particularly given that calculations must be performed over a grid of collision energies. Moreover, even with such a large number of channels, we were not able to obtain the converged value of the pressure shift cross-section or the imaginary part of the Dicke cross-section. This could be caused by the fact that we are using a simplified version of the HCN-N$_{2}$ potential energy surface, and both cross-sections are particularly susceptible to the fine details of the PES. Hence, neither the pressure shift nor the imaginary part of the Dicke coefficient were calculated in this work.

Given the computational infeasibility of fully converged calculations, an alternative approach was necessary. We explored whether a simplified basis including only asymptotically open channels could provide sufficiently accurate results. Remarkably, for the test case of the R(0) line with $j_{2}=0$ at $E_{\mathrm{kin}}=100$~cm$^{-1}$, using only open channels led to real parts of the generalized cross-sections $\sigma^{\kappa}_{\lambda}$ that differed by merely 4\% from the reference $j_{1}^{\mathrm{max}}=15$ and $j_{2}^{\mathrm{max}}=14$ case, for both 
$\lambda=0$ and $\lambda=1$.

How is this possible? Recall that one of the key factors determining the relevance of a particular scattering channel is the strength of its coupling to the channels of interest. Let us examine this case in detail.

For the R(0) line with $j_{2}=0$, the primary scattering channels of interest (see Eq.~\eqref{eq:crosssection}) asymptotically converge to two degenerate sets: (1) channels with $j_{a}=0$, $j_{2}=0$, $j_{a2}=0$ and all possible values of $L$ within a given $\mathcal{J}_{a}$ block, (2) channels with $j_{b}=1$, $j_{2}=0$, $j_{b2}=1$, and all possible $\bar{L}$ values within a given $\mathcal{J}_{b}$ block.

At a collision energy of $100$~cm$^{-1}$, a large number of channels are energetically accessible. Focusing on the $j_{a}=0$, $j_{2}=0$ case, this includes 425 channels, formed from combinations of 24 different $j_{1}, j_{2}$ pairs with $E_{j_{1}}+E_{j_{2}} < 100$~cm$^{-1}$. Our understanding is that this reduced set of open channels already contains those that are most strongly coupled via the dominant anisotropic terms in the PES expansion (Eq.~\eqref{fit}), $v_{l_1 l_2 \mu}(R)$. These terms are responsible for redistributing scattering amplitude among different channels, and their inclusion appears to be sufficient to achieve a few-percent convergence of the cross-sections.

This finding presents a clear choice: we could either attempt fully converged calculations, which remain computationally prohibitive while providing only marginal improvements in accuracy, or adopt the open channels approach, which retains a few-percent accuracy while dramatically reducing computational cost. For the tested case, the number of energetically accessible channels never exceeded 425, which is nearly a factor of 36 smaller than the staggering number of 15~000 channels included in the reference $j_{1}^{\mathrm{max}}=15$, $j_{2}^{\mathrm{max}}=14$ case. Since computational time scales with the number of channels $N$ as $N^{3}$~\citep{Truhlar_1994}, this reduction translates into an estimated speedup of four orders of magnitude, making large-scale calculations in the HCN-N$_{2}$ system feasible.

The open channels approach, as described above, inherently requires the basis set to be dynamically adjusted with collision energy. Since only energetically accessible channels are included, the number of channels naturally increases with increasing collision energy. This means that the method should, in principle, perform better at higher collision energies, as more scattering channels---coupled through different anisotropic terms---are naturally incorporated into the calculations.

However, at sufficiently low collision energies, where only a few channels remain open, the exclusion of closed channels is expected to be more significant. While our test case suggested that dominant coupling effects were well captured within the open channel space, the validity of this approximation in the cold regime ($E_{\mathrm{kin}}\sim10^{1}$~cm$^{-1}$) requires further study. Likewise, at very high collision energies, the approach eventually reaches a computational ceiling, as the number of open channels may push the limits of feasibility, making additional refinements less practical.

A detailed investigation of the open channels method, including its applicability across different collision energies and molecular systems, as well as its comparison with fully converged quantum scattering calculations, is currently underway. In this work, we adopt this method as it provides the best balance between computational feasibility and accuracy, and we validate these approximate calculations by a direct test against experimental line shape parameters for N$_{2}$-perturbed HCN lines.

\section{Results and discussion}
\label{sec4}
Table~\ref{tab:coefficients} summarizes the experimental results of \ce{N2} pressure broadening, its speed-dependence, and pressure shift coefficients of the R(0), R(1), R(2) transitions of \ce{HCN}. Moreover, the theoretical estimates of the pressure broadening and its speed-dependence coefficients of the lowest five rotational transitions are reported, along with the comparison with their experimental counterparts (when available).   
\begin{table}[b!]
    \centering
    \renewcommand{\arraystretch}{1.2}
    \caption{Experimental values (Exp.), theoretical estimates (Theo.) and percentage deviations (\% Dev.) of the \ce{N2} pressure broadening ($\gamma_0$) and speed-dependent relaxation ($\gamma_2$) coefficients for the R(0), R(1) and R(2) transitions of \ce{HCN} at 296\,K, which extend up to R(4) for the computed ones. Experimental pressure shift coefficients ($\delta_0$) for the R(0), R(1) and R(2) transitions are also reported.}
    \label{tab:coefficients}
    \footnotesize
    \begin{tabular}{c ccc ccc c}
        \toprule
         & \multicolumn{3}{c}{$\gamma_0$} & \multicolumn{3}{c}{$\gamma_2$} & $\delta_0$ \\
        \midrule
        & \multicolumn{3}{c}{($10^{-3}\times$ cm$^{-1}$/atm)\textsuperscript{b}} & \multicolumn{3}{c}{($10^{-3}\times$cm$^{-1}$/atm)} & \multicolumn{1}{c}{($10^{-3}\times$cm$^{-1}$/atm)} \\
        \cmidrule(r){2-4} \cmidrule(l){5-7}
        & Exp. & Theo.\textsuperscript{a} & \% Dev. & Exp. & Theo.\textsuperscript{a} & \% Dev. & Exp. \\
        \midrule
R(0) & 167.08(44) & 161.4 & 3.4 & 15.89(23)  & 14.1  & 11.2 & -3.01(86) \\
R(1) & 172.17(81) & 160.0 & 7.1 & 31.78(76)  & 20.6  & 35.2 &  -0.8(11)  \\
R(2) & 160.51(61) & 157.8 & 1.7 & 26.59(56)  & 26.2  & 1.5  &  -0.3(22)  \\
R(3) &               & 148.0 &     &               & 26.5  &     & \\
R(4) &               & 140.7 &     &               & 28.5  &     & \\

        \bottomrule
    \end{tabular}
    \vspace{2mm}
    
    \textsuperscript{a} See \S~\ref{sub3.2} for details. \\
    \textsuperscript{b} To maintain consistency with HITRAN notation, in the following, $10^{-3}\times$ cm$^{-1}$/atm units will be adopted. Conversion factor: [$10^{-3}\times$ cm$^{-1}$/atm] $\times \, 0.039446 =$ [MHz/Torr].
\end{table}
In general, there is a good agreement between the computed and experimental values, although the open channels approach tends to underestimate both the $\gamma_0$ and $\gamma_2$ coefficients. This result is particularly promising, especially considering that the decision to neglect the $l_2>2$ terms in the fitting of the potential prevents the correct coupling of \ce{N2} energy levels with $\Delta J (\ce{N2})> 2$. This further supports the accuracy of the open channels approach used for scattering calculations.
The largest deviations are observed in the R(1) transition of \ce{HCN}. This was somehow expected, as part of the information of the most intense hyperfine transitions is not included in the central line. Additionally, the hyperfine interaction may have introduced subtle distortions in the line shape, leading to an increased effective broadening difficult to model from a theoretical point of view.
As $J$ increases, the dominant character of the hyperfine transitions progressively concentrates in the central line and the hyperfine coupling becomes weaker. Thus, these discrepancies are expected to decrease with increasing $J$.
Within the pressure range explored in this study, the collision shift was very small and could not be reliably determined due to the presence of comparable if not larger systematic uncertainties. In the most extreme case, for the R(2) transition, the collision shift coefficient was barely determined. 
Given the uncertainties involved in the experimental determination of the pressure shift and the limited reliability of the open channels approach in its prediction, a comparison with computed values was not considered meaningful.

\begin{figure*}  
\centering
 \includegraphics[scale=0.71]{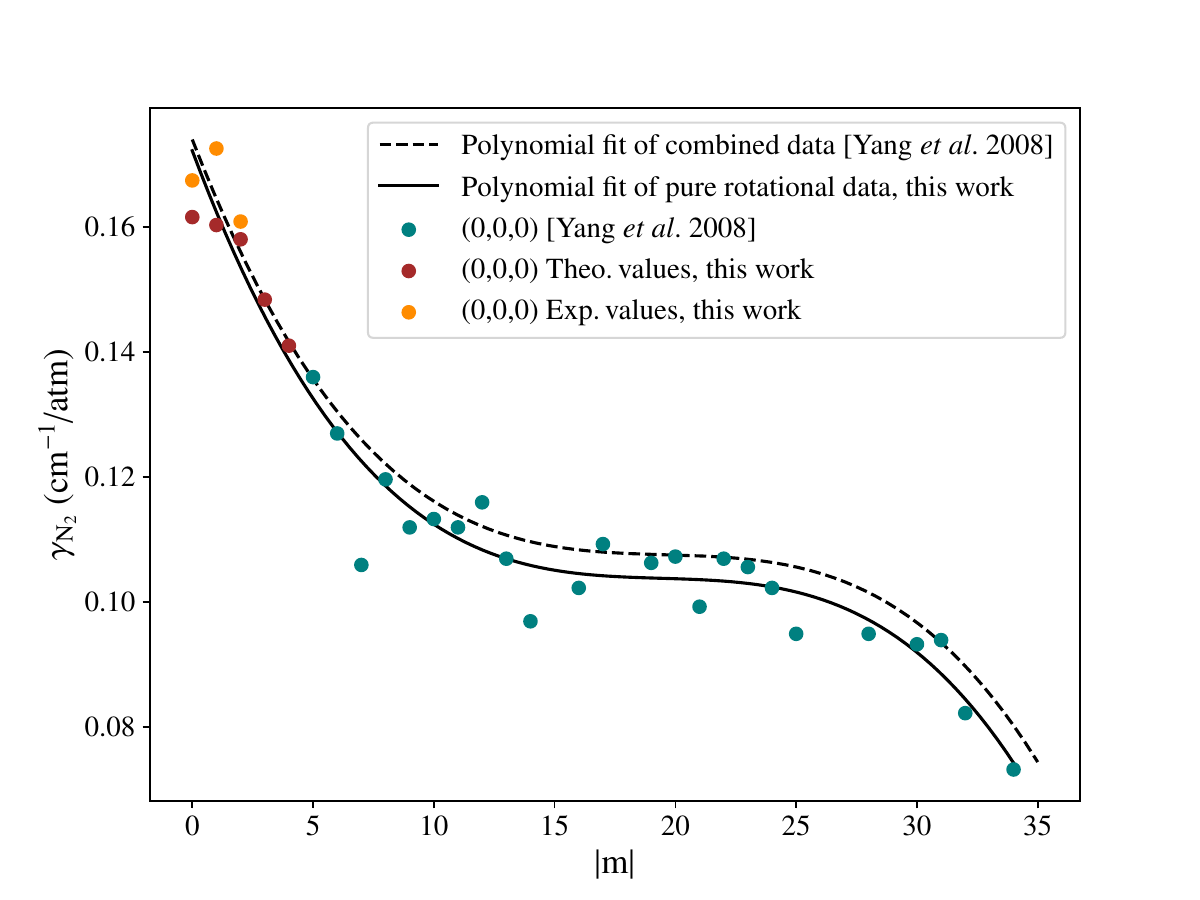}
 \caption{Comparison of the predictions made by third-order polynomial fit of the purely rotational pressure broadening values at 296\,K determined in this work and in~\citet{yang2008oxygen} with the ones made by~\citet{yang2008oxygen} on a combined set of data with the vibrational and isotopically substituted counterparts.}
 \label{fig7} 
\end{figure*}
In~\citet{yang2008oxygen}, the comparison of nitrogen-broadened half-widths for the R-branch lines of \ce{HCN} in different vibrational states suggested a possible vibrational dependence of these coefficients. The extension of the experimental dataset in this work allows for further investigation of this comparison through a polynomial expansion that uniquely characterizes the collisional broadening of purely rotational lines. 
Specifically, the \ce{N2} broadening coefficients were estimated using the third-order polynomial expression employed by~\citet{yang2008oxygen}:
\begin{equation}
 \gamma_0=A_0+A_1|m|+A_2|m|^2+A_3|m|^3\,,
\label{eq1}    
\end{equation}
where $m=-J''$ for the P-branches and $m=J'$ for the R-branches of \ce{HCN}.
The results of this analysis are shown in Fig.~\ref{fig7}, where the purely rotational pressure broadening coefficients are plotted against $m$. The polynomial fit was then compared to the polynomial fit obtained by~\citet{yang2008oxygen}, which includes values for vibrationally excited states and the \ce{HC^{15}N} isotopologue. 
The comparison of polynomial coefficients, presented in Table~\ref{tab:polynomial}, revealed non-negligible differences. For an accurate prediction of the \ce{N2} pressure broadening coefficients of purely rotational \ce{HCN} transitions, the polynomial fit derived in this work is thus recommended. 
\begin{table}[t!]
    \centering
    \renewcommand{\arraystretch}{1.2}
    \caption{Polynomial fits of \ce{N2} broadened half-widths of \ce{HCN} at 296\,K.}
    \label{tab:polynomial}
    \footnotesize
    \begin{tabular}{c cc c cc}
        \toprule
        & \multicolumn{2}{c}{Pure rotational lines} & &\multicolumn{2}{c}{All bands from~\citet{yang2008oxygen}} \\
        \cmidrule(r){2-3} \cmidrule(r){5-6}
        Parameter\textsuperscript{a} & Value\textsuperscript{b} & Error & & Value\textsuperscript{b} & Error \\
        \midrule
        $A_0$ & \phantom{-}0.171\phantom{$\times10^{-2}$} & 2.7$\times10^{-3}$  && \phantom{-}0.174\phantom{$\times10^{-2}$} & 1.1$\times10^{-3}$ \\
        $A_1$ & -1.030$\times10^{-2}$ & 8.3$\times10^{-4}$ && -0.997$\times10^{-2}$ & 2.7$\times10^{-4}$  \\
        $A_2$ & \phantom{-}5.271$\times10^{-4}$ & 6.2$\times10^{-5}$ &&  \phantom{-}5.039$\times10^{-4}$ & 1.8$\times10^{-5}$  \\
        $A_3$ & -9.844$\times10^{-6}$ & 1.2$\times10^{-6}$ && -8.582$\times10^{-6}$ & 3.5$\times10^{-7}$  \\
        $R^2$ & \multicolumn{2}{c}{0.96} && \multicolumn{2}{c}{0.97} \\
        \bottomrule
    \end{tabular}
    \vspace{2mm}  
    
    \textsuperscript{a} Parameters defined according to the polynomial expression of Eq.~\eqref{eq1}. \\
    \textsuperscript{b} Broadening coefficients are in cm$^{-1}$/atm.
\end{table}

The good agreement shown by the comparison of the experimental values of the $\gamma_0$ and $\gamma_2$ coefficients with their calculated counterparts confirmed the accuracy of the theoretical strategy adopted in this work. Therefore, in order to ensure a proper applicability of these coefficients for the modeling of both Titan's and terrestrial atmospheres, the estimated theoretical values were extended to a wide range of temperatures. This has also been done for the computational estimate of the Dicke narrowing coefficient, which is difficult to determine by experimental means. The three coefficients of the R(0), R(1), R(2), R(3) and R(4) transitions were thus computed over a temperature range from 100 to 800\,K. The results were then fitted to single power-law distributions 
according to the following expressions:
\begin{equation}
\label{SPL_fit}
    \begin{aligned}
        \gamma_0(\text{T}) &= g_0 (\text{T}_{\text{ref}}/\text{T})^n\,,\\
        \gamma_2(\text{T}) &= g_2 (\text{T}_{\text{ref}}/\text{T})^j\,,\\
        \tilde\nu_{opt}^r(\text{T}) &= r (\text{T}_{\text{ref}}/\text{T})^p\,,\\
    \end{aligned}
\end{equation}
where T$_{\text{ref}}=296$\,K.
The resulting temperature dependence coefficients are summarized in Table~\ref{tab:fit_singlepowerlaw}. 
\begin{table}[h!]
    \centering
    \renewcommand{\arraystretch}{1.2}
    \caption{Temperature dependence coefficients obtained from the single power-law fit on the theoretically computed values\,\textsuperscript{a}.}
    \label{tab:fit_singlepowerlaw}
    \footnotesize
    \begin{tabular}{c cc c cc c cc}
        \toprule
        & \multicolumn{2}{c}{$\gamma_0$(T)} & & \multicolumn{2}{c}{$\gamma_2$(T)} & & \multicolumn{2}{c}{$\tilde\nu^r_{\text{opt}}$(T)} \\
        \cmidrule(r){2-3} \cmidrule(r){5-6} \cmidrule(r){8-9}
         & $g_0$\textsuperscript{b} & $n$ & & $g_2$\textsuperscript{b} & $j$ & & $r$\textsuperscript{b} & $p$ \\
        \midrule
        R(0)  & 0.161 & 0.817 && 0.014 & 1.030 && 0.011 & 0.967 \\
        R(1)  &  0.160 &  0.718 &&  0.021 &  0.853 &&  0.017 &  0.820 \\
        R(2) & 0.159 & 0.623 && 0.027 & 0.658 && 0.022 & 0.660 \\
        R(3) & 0.149 & 0.605 && 0.027 & 0.628 && 0.026 & 0.549 \\
        R(4) & 0.142 & 0.555 && 0.029 & 0.580 && 0.025 & 0.573 \\
        \bottomrule
    \end{tabular}
    \vspace{2mm}  
    
    \textsuperscript{a} Parameters defined according to the expressions of Eq.~\eqref{SPL_fit}. \\
    \textsuperscript{b} $g_0$, $g_2$ and $r$ coefficients are in cm$^{-1}$/atm. \\
    \textsuperscript{c} The expected uncertainties are in the order of $5-10$ \%.
\end{table}

\section{Conclusions}
\label{sec5}
Having precise knowledge of pressure-induced coefficients is essential for accurately modeling the rotational lines of molecules relevant to both terrestrial and astrophysical studies. This is the case of \ce{HCN}, a molecule of significant importance in atmospheric monitoring, climate evolution simulations,
and radiative transfer modeling in both terrestrial and astrophysical environments. 

However, despite numerous theoretical and experimental investigations of the effect of perturber gases on the ro-vibrational lines of \ce{HCN}, there remains a complete lack of data on its broadening for low-energy, purely rotational transitions caused by \ce{N2}. Currently, the HITRAN database provides broadening values for transitions involving $J(\ce{HCN})<5$ by extrapolating from a global fit that incorporates only purely rotational values for higher $J(\ce{HCN})$, and assuming no dependence on the vibrational state or isotopic species of \ce{HCN}. Moreover, there is no available data on \ce{HCN} coefficients beyond the standard Voigt profile --- e.g., speed-dependent broadening coefficient or the Dicke narrowing caused by velocity-changing collisions. 

In this work, we addressed this lack by experimentally determining room-temperature \ce{N2} pressure broadening coefficients, their speed-dependencies, and pressure shift coefficients for the three lowest rotational transitions of \ce{HCN}. 

Our experimental results were also used to validate a novel computational strategy based on open channels scattering calculations, and employing a simplified potential representation that considers only five orientations of the perturber. This approach aimed at remarkably reducing the computational eﬀort without compromising the accuracy.
The computational-experimental comparison showed a good agreement, with average percentage deviations of approximately $4\%$ for pressure broadening coefficients and less than $\sim 16\%$ for speed-dependent broadening values. 
Building on this validation, the same computational strategy was applied to compute the \ce{N2} pressure broadening coefficients, their speed-dependencies, and Dicke narrowing coefficients, as well as their temperature dependence coefficients --- fitted $via$ single power-law distribution for a dataset between 100 and 800\,K --- for all the rotational transitions of \ce{HCN} with $J<5$. 
The newly obtained experimental and computational pressure broadening coefficients were then used to refine the polynomial fit employed in the HITRAN database, this time considering only purely rotational data for $0<J(\ce{HCN})<34$. These results are pivotal to support and improve the modeling of \ce{HCN} rotational lines observed in both the terrestrial and Titan's atmospheres. Beyond the improvements to the HITRAN database, this study also lays the groundwork for calculating state-to-state HCN-N$_{2}$ rate coefficients, which will be valuable for non-LTE simulations of astrophysical environments. Work in this direction is currently underway.

\paragraph{Funding:} 
This work was supported by MUR (PRIN Grant Numbers 20\-20\-82\-CE\-3T, P2022ZFNBL and 20225228K5) and by the University of Bologna (RFO funds). 
The COST Action CA21101 ``COSY - Confined molecular systems: from a new generation of materials to the stars’’ is also acknowledged.
The project was also supported by the National Science Centre in Poland through Project Nos. 2022/46/E/ST2/00282 (P.W) and 2019/35/\-B/\-ST2\-/\-01118 (H.J). For the purpose of Open Access, the authors have applied a CC-BY public copyright licence to any Author Accepted Manuscript (AAM) version arising from this submission. We gratefully acknowledge Polish high-performance computing infrastructure PLGrid (HPC Center: ACK Cyfronet AGH) for providing computer facilities and support within computational grant no. PLG/2024/017376. Created using resources provided by Wroclaw Centre for Networking and Supercomputing (\url{http://wcss.pl)}. We also acknowledge Rennes Metropole for financial support.

%

\clearpage

\bibliographystyle{elsarticle-num-names} 
\bibliography{bibliography.bib}

\end{document}